\documentclass[fleqn,usenatbib]{mnras}

\usepackage{newtxtext,newtxmath}
\usepackage[T1]{fontenc}
\usepackage{ae,aecompl}
\usepackage{graphicx}	% Including figure files
\usepackage{amsmath}	% Advanced maths commands
\usepackage{amssymb}	% Extra maths symbols

\title[Stellar Feedback in NGC~3351]{Survival of molecular gas in a stellar feedback-driven outflow witnessed with the MUSE TIMER project and ALMA}
\author[R. Leaman et al.]{
Ryan Leaman$^{1}$\thanks{E-mail: leaman@mpia.de},
Francesca Fragkoudi$^{2}$,
Miguel Querejeta$^{3,4}$,
Gigi Y.C. Leung$^{1}$,
\newauthor
Dimitri A. Gadotti$^{3,5}$,
Bernd Husemann$^{1}$,
Jesus Falc\'{o}n-Barroso$^{6,7}$,
\newauthor
Patricia S\'{a}nchez-Bl\'{a}zquez$^{8}$,
Glenn van de Ven$^{3,9}$,
Taehyun Kim$^{10}$,
Paula Coelho$^{11}$,
\newauthor
Mariya Lyubenova$^{3}$,
Adriana de Lorenzo-C\'{a}ceres$^{6,7}$,
Marie Martig$^{12,1}$,
\newauthor
Inma Martinez-Valpuesta$^{6,7}$,
Justus Neumann$^{5,13}$,
Isabel P\'{e}rez$^{14,15}$
and Marja Seidel$^{16}$
\\
$^{1}$Max-Planck Institut f\"ur Astronomie, K\"onigstuhl 17, D-69117 Heidelberg, Germany\\
$^{2}$Max-Planck-Institut f\"{u}r Astrophysik, Karl-Schwarzschild-Str. 1, 85741 Garching, Germany\\
$^{3}$European Southern Observatory, Karl-Schwarzschild-Str. 2, D-85741 Garching, Germany\\
$^{4}$Observatorio Astron{\'o}mico Nacional (IGN), C/Alfonso XII 3, Madrid E-28014, Spain\\
$^{5}$European Southern Observatory, Ave. Alonso de Cordova 3107, Casilla 19, 19001, Santiago, Chile\\
$^{6}$Instituto de Astrof\'{i}sica de Canarias, Calle V\'{i}a L\'{a}ctea s/n, E-38205 La Laguna, Tenerife, Spain\\
$^{7}$Departamento de Astrof\'{i}sica, Universidad de La Laguna, E-38200 La Laguna, Tenerife, Spain\\
$^{8}$Departamento de F\'{i}sica Te\'{o}rica, Universidad Aut\'{o}noma de Madrid, E-28049 Cantoblanco, Spain\\
$^{9}$Department of Astrophysics, University Vienna, T\"{u}rkenschanzstrasse 17, 1180 Wien, Austria\\
$^{10}$Korea Astronomy and Space Science Institute, Daejeon, 34055, Korea\\
$^{11}$Instituto de Astronomia, Geof\'{i}sica e Ci\^{e}ncias Atmosf\'{e}ricas, Universidade de S\~{a}o Paulo, R. do Mat\~{a}o 1226, 05508-090 S\~{a}o Paulo, Brazil\\
$^{12}$Astrophysics Research Institute, Liverpool John Moores University, Liverpool Science Park ic2, 146 Brownlow Hill, Liverpool L3 5RF, UK\\
$^{13}$Leibniz-Institut f\"{u}r Astrophysik Potsdam (AIP), An der Sternwarte 16, D-14480 Potsdam, Germany\\
$^{14}$Departamento de F\'{i}sica Te\'{o}rica y del Cosmos, Universidad de Granada, Campus de Fuentenueva, E-18071 Granada, Spain\\
$^{15}$Instituto Carlos I de F\'{i}sica Te\'{o}rica y Computacional, Universidad de Granada, E-18071 Granada, Spain\\
$^{16}$Observatories of the Carnegie Institution of Washington, Pasadena, CA 91101, USA
}

\date{Accepted XXX. Received YYY; in original form ZZZ}

\pubyear{2015}

\begin{document}
\label{firstpage}
\pagerange{\pageref{firstpage}--\pageref{lastpage}}
\maketitle

\begin{abstract}
Stellar feedback plays a significant role in modulating star formation, redistributing metals, and shaping the baryonic and dark structure of galaxies -- however, the efficiency of its energy deposition to the interstellar medium is challenging to constrain observationally. Here we leverage HST and ALMA imaging of a molecular gas and dust shell ($M_{H2} \sim 2\times 10^{5} ~{\rm M}_{\odot}$) in an outflow from the nuclear star forming ring of the galaxy NGC 3351, to serve as a boundary condition for a dynamical and energetic analysis of the outflowing ionised gas seen in our MUSE TIMER survey.  
%While the total energy of this stellar feedback-driven outflow is comparable to low luminosity AGN, 
We use \texttt{STARBURST99} models and prescriptions for feedback from simulations to demonstrate that the observed star formation energetics can reproduce the ionised and molecular gas dynamics -- provided a dominant component of the momentum injection comes from direct photon pressure from young stars, on top of supernovae, photoionisation heating and stellar winds. The mechanical energy budget from these sources is comparable to low luminosity AGN, suggesting that stellar feedback can be a relevant driver of bulk gas motions in galaxy centres - although here $\lesssim 10^{-3}$ of the ionized gas mass is escaping the galaxy.  
%Given the short expansion distance for the outflow, we rule out in-situ condensation as an origin for the molecular gas at the leading edge of the wind.  The morphology of the CO shell and emission line diagnostics instead suggest a scenario where magnetic field lines in the dusty ISM aided its survival as it was launched from the ring by stellar radiation pressure.  The magnetic field strengths required, are in excellent agreement with analytic disk equilibrium models and geometry of the dust shell.
We test several scenarios for the survival/formation of the cold gas in the outflow, including in-situ condensation and cooling.  Interestingly, the geometry of the molecular gas shell, observed magnetic field strengths and emission line diagnostics are consistent with a scenario where magnetic field lines aided survival of the dusty ISM as it was initially launched (with mass loading factor $\lesssim 1$) from the ring by stellar feedback.
%MCMC analysis of these data and analytic feedback models suggests that the photon trapping factor used in hydrodynamical galaxy simulations which best reproduces the observed gas configuration corresponds to $\kappa_{IR} = 44\pm 27 ~{\rm cm}^{2}~ {\rm g}^{-1}$, with a shell expansion time of $\sim 6$ Myrs and initial mass loading factor of $\eta = 0.1-1$.  
This system's unique feedback driven morphology can hopefully serve as a useful litmus test for feedback prescriptions in magnetohydrodynamical galaxy simulations.      %We also show how the maximal radial extent of the dusty gas shells provides an estimate of the burstiness of the recent SF, finding $\Delta t_{burst} \sim 150$ Myr, while their angular separation in barred galaxies further provides a novel method of estimating the bar speed - producing comparable results to classic continuity methods, $\Omega_{bar} = 37 \pm 5$ km s$^{-1}$ kpc$^{-1}$.
\end{abstract}

\begin{keywords}
galaxies: evolution -- galaxies: kinematics and dynamics  --  galaxies: ISM -- ISM: kinematics and dynamics -- galaxies: magnetic fields
\end{keywords}

%%%%%%%%%%%%%%%%%%%%%%%%%%%%%%%%%%%%%%%%%%%%%%%%%%

%%%%%%%%%%%%%%%%% BODY OF PAPER %%%%%%%%%%%%%%%%%%

\section{Introduction}
Feedback from young stars throughout their lifetime and at their death is thought to be a significant contributor to the energy budget of the interstellar medium (ISM) within star forming galaxies.  Stellar evolution theory has provided many quantitative predictions for how feedback (stellar winds, ionizing radiation, metals and supernovae (SNe)) vary as a function of time, metallicity and initial stellar mass function (IMF) within a population of stars (c.f., \citealt{Leitherer99}).  However, the subsequent impact of these events on the host galaxy have been characterized best through numerical hydrodynamic simulations of galaxies.

%Role/consequences of feedback: core/cusp, metal loss, SF suppression, outflows AND Observational signatures of feedback: outflows, AGN, CGM metals, shocks\\
The inclusion of baryonic physics within galaxy simulations has helped solve a plethora of discrepancies between dark matter (DM) only simulations and observed galaxy properties \citep{Katz92,Navarro93}. Newer implementations of feedback (e.g., \citealt{Stinson06,Hopkins14}) proved to be a crucial ingredient in removing low angular momentum gas within galaxies -- helping simulations form realistic low bulge fractions and radially extended thin galactic disks \citep{Governato04}.  In addition, SNe feedback has been shown to drive metals into the circumgalactic media via outflows (e.g., \citealt{Shen13,Turner16}), which are observed commonly around galaxies of many types and masses (e.g., \citealt{Werk16}) -- and this process is also thought to modify the cuspy DM profiles in low mass galaxies \citep{Dicintio14}.
%
%The role of stellar feedback in suppressing star formation (SF) in low mass galaxies is also thought to be important for matching observed star formation histories (SFH; \citealt{Zolotov12,Cole14}), reproducing the stellar mass metallicity relation \citep{Gallazzi05,Brooks07,Kirby11} and suppressing stellar mass growth in order to reproduce stellar-to-halo dark matter relations (from abundance matching or satellite mass functions around the Milky Way; \citealt{Sawala15}). 

%Modern simulation parameterizations\\
The ever improving numerical resolution among hydrodynamical simulations has opened up a wealth of new investigative possibilities, but also added complexity in how sub-grid feedback prescriptions are modeled.  Such prescriptions may try to address for example: the efficiency of photon pressure on dust \citep{Hopkins11}, thermalization of the energy dumped into the ISM by feedback, or porosity/escape fraction effects \citep{KrumThomp13}. Recent studies have also focused on the role played by ionizing radiation from young massive stars -- as the energy budget could be comparable to the SNe of a given simple stellar population (SSP) in some cases (e.g., \citealt{Rosdahl15}).

Reproduction of observed galaxy mass functions, and even small scale dynamical, chemical and structural features \citep{, Few12,Schroyen13,Renaud13,Martig14,Dutton16,Elbadry16,Read16,Dicintio17} is a great success for these simulations and models. 
%In particular feedback driven core creation \citep{Governato10} in the dark matter and baryonic component of low mass galaxies has led to recent success in reproducing the inner rotation curves and extended sizes of low mass dwarf galaxies \citep{Dicintio17}.  
However, concerns about the degenerate behaviour of many of the complex prescriptions used for feedback or star formation (e.g., \citealt{CloetOsselaer12}) provides ample motivation for the observational community to continue to provide useful constraints for calibration of these methods.
 
%Observational tests of subgrid prescriptions\\
Direct observations of large scale outflows offer a natural window into the energetics provided by feedback sources (e.g., \citealt{Martin05,Lopez11}).  Finding a boundary condition for the initial time or total distance of the outflow event is often observationally problematic to establish, and prevents a complete evolutionary picture and energetic accounting of galactic systems.  Nevertheless instantaneous observational quantities such as the outflow mass loading factor ($\eta = \dot{M}_{\rm out} / SFR$) are essential to calibrate energy and momentum deposition in simulations \citep{Dave11,Schroetter16}. 
Additionally several indirect tests of the impact of feedback processes have been performed at high redshift, and in the local universe, such as exploring the radial metallicity profiles of disk galaxies, which were shown to be sensitive to the feedback prescriptions used in simulations (e.g., \citealt{Pilkington12,Gibson13,Schroyen13,Elbadry16,Starkenburg17}). 

%Studying feedback at intermediate scales, where the observational resolution is comparable to those of simulations, is limited to the small number of well studied examples in the Local Universe.
Motivated by the ambitions of modern galaxy simulations and past observational tests, in this paper we explore stellar feedback at intermediate scales, where the observational resolution is comparable to that of current galaxy simulations. We present kinematic observations of the multi-phase ISM in the nuclear ring galaxy NGC~3351, as part of our VLT/MUSE TIMER survey (PI: Gadotti)\footnote{https://www.muse-timer.org/}, with the goal of understanding the survival of cold gas in outflows and providing quantitative tests for sub-grid stellar feedback prescriptions.

\begin{figure*}
\begin{center}
\includegraphics[width=0.94\textwidth,angle=0]{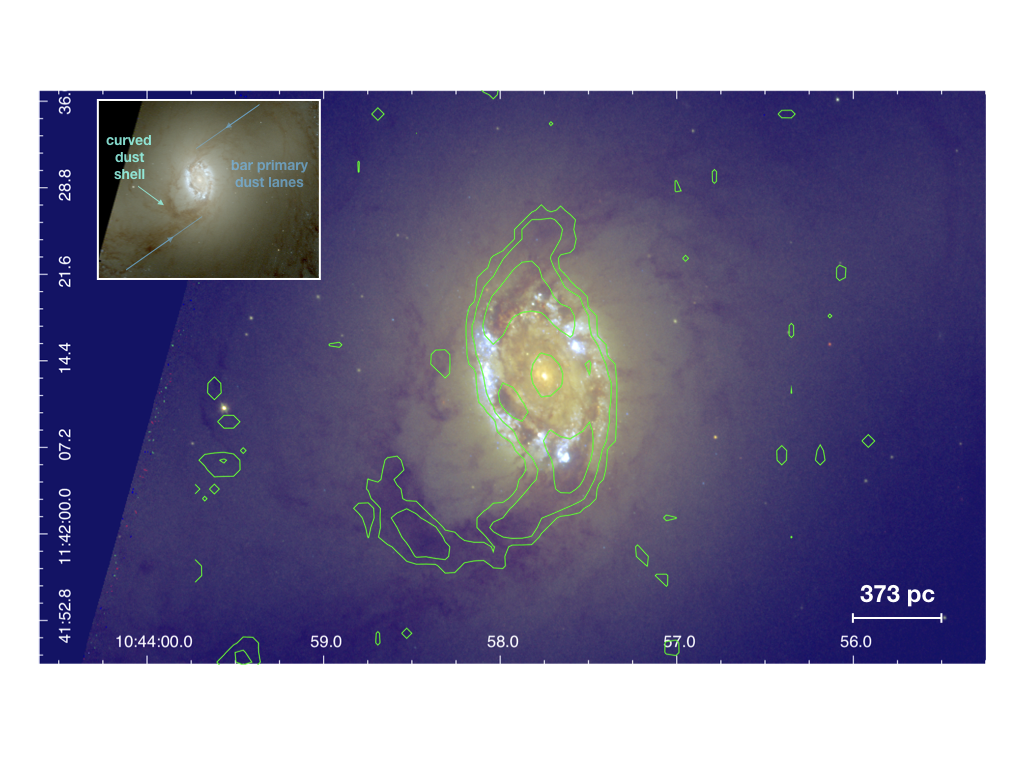}
\caption{HST WFC3 colour composite image of the central region of NGC~3351 made as part of the LEGUS survey (https://legus.stsci.edu/) from F814W, F555W and F336W exposures.  In the inset, we show a zoomed-out image of the galaxy, where we can see the `primary dust lanes' caused by bar-driven shocks (their length and orientation are indicated with the blue lines); the elliptical dust shell (which is pointed to by the arrow) is evident as a curved transverse feature,  orthogonal to the `primary' feeding lanes of the bar. This primary shell is coincident with molecular gas emission in the ALMA moment 0 CO map (\emph{green contours}).}
\label{fig:hst3col}
\end{center}
\end{figure*}

NGC~3351 is comparable to our own Galaxy in total stellar mass, metallicity ($M_{*} = 3.1\times10^{10}~{\rm M}_{\odot}$, $Z/Z_{\odot} = 0.4$) and star formation (SFR$ = 0.6~ {\rm M}_{\odot} {\rm yr}^{-1}$ in the nuclear ring alone; \citealt{Gadotti19}). As with all of the galaxies in the TIMER survey, it was selected because of its strongly barred morphology and presence  of a nuclear ring - with its classification listed as $(R^{'})SB(r,bl,nr)a$ in \citep{Buta15}. NGC 3351 exhibits dust lanes throughout the spiral arms, along the bar and nuclear ring - along with a transverse dust lane offset from the nuclear ring and orthogonal to the bar lanes.  This feature will be discussed throughout this work and is indicated by the arrow in Figure \ref{fig:hst3col}. NGC~3351 is considered an isolated field galaxy \citep{Young96} and NED lists the closest neighbour (Messier 96) as having projected distance and redshift separations of $D_{sep} = 125$ kpc and $\Delta_{cz} = 120$ km s$^{-1}$ respectively (\url{https://ned.ipac.caltech.edu/forms/denv.html}) -  implying minimum interaction timescales of at least 1000 times longer than the dynamical time of the nuclear ring in NGC~3351. While this is a heterogeneous source catalogue of detections, the list appears luminosity complete down to values of $ L \sim 4\times10^{5} L_{\odot}$.  Notably, NGC~3351 has no evidence for an AGN (from emission line diagnostics; e.g., \citealt{Gadotti19}) and as such is often used as a control galaxy in studies on AGN properties and their feeding mechanisms (e.g., \citealt{Colina97}). 

The MUSE observations presented here show a serendipitous and visually clear signature of a planar/radial outflow from the nuclear ring of NGC~3351 in the ionised gas flux and kinematic maps.  This was accompanied by identification of a unique concentric shell of dust and molecular gas in archival HST and ALMA observations. 
The location of this transverse dusty gas shell is unexpected from a purely gravitational standpoint (see Sect. 3), which teases at a common evolution with the ionised gas, likely due to feedback from the  star-forming nuclear ring.
In addition, as seen and discussed in other galactic outflows, the presence of cold molecular gas may be naively unexpected given the hot, energetic phenomena that provide kinematic feedback in the ISM (c.f., \citealt{Mccourt18}).  This system offers a relatively complete observational framework to test many of the theoretical models proposed for cold gas formation and survival in outflows.

\subsection{A Stellar Feedback Scenario for the ISM in NGC~3351}
%Our observations show that unlike most typical nuclear ring galaxies, there is substantial gas and cold dust in a semi-elliptical transverse shell, concentric to the nuclear SF ring, and orthogonal to the ubiquitous linear dust/gas feeding lanes which contact the north and south edges of the nuclear ring.  This feature is unexpected - as unless significant gas instabilities are present, purely dynamical forces effectively prevent gas from forming in large quantities in this region, as shown above.  We present below a \textit{stellar feedback based scenario for the origin of this feature}.

The clear signatures of warm gaseous outflow in our data, together with the archival ALMA and HST data of the cold gas and dust, hint at a scenario for the transverse dust-lane feature:  (1) stellar feedback from the nuclear ring pushes out the cold gas shell, which was initially co-located with the nuclear SF ring, thus forming a low density cavity. (2) Warm ionised gaseous outflows, which fill the cavity, shock with the underlying multi-phase ISM. (3) The survival of the cold phase during the outflow could have been aided by magnetic fields initially permeating through the dusty ISM, which is consistent with a vertically constrained outflow for the hot ionized gas.

While in principle other scenarios such as cold gas inflows or fly-by with a satellite galaxy, could potentially be responsible for the origin of the curved dust lane observed in NGC\,3351, we discuss in Section \ref{sec:origin} why these are less likely than the feedback scenario proposed here - and note the isolation criteria mentioned above rule out a fly-by origin for the feature given the relevant timescales.
In what follows, we quantitatively evaluate the validity of this feedback scenario and scenarios for the survival of the cold gas, in the context of the data, stellar feedback models, and numerical simulations.
The observations are presented in Sect. 2 of this paper. In Sect. 3 we present hydrodynamical simulations tailored to reproduce the effect of the gravitational potential of NGC 3351 on the gas-dynamics. We analyze the emission-line and dynamical properties of the ionized and molecular gas in Sect. 4. We present a model for the feedback processes and dynamics of the outflow observed in this system in Sect. 5, and discuss the survival of molecular gas in outflows in Sect. 6. Implications of our findings for feedback parametrizations in simulations and further astrophysical questions are discussed in Sect. 7.  We summarize this study with our main conclusions in Sect. 8.  Where necessary we adopt a Hubble constant of $H_{0} = 67.8$ km s$^{-1}$ Mpc$^{-1}$ and $\Omega_{m} = 0.308$ with a flat topology (Planck Collaboration et al., 2015).

\section{Observations and Analysis}
As much of the analysis in this work surrounds the dynamics, stellar populations and gas content of the inner region of NGC~3351, we briefly describe some of the relevant features and their properties for context.  Figure \ref{fig:hst3col} highlights several of the structures of interest in our study.  The nuclear ring is evident in molecular gas and exhibits ongoing star formation typical of these structures.  In NGC~3351, the nuclear ring has a radius of $\sim 250$ pc and width of $\sim 50-100$ pc.  The bar of the galaxy, which can be traced in the image through the linear dust feeding lanes, is orientated 112 degrees from North \citep{Salo15}, and has a semi-major axis of 2.7 kpc \citep{Kim14}.  Given the derived bar mass fraction from photometric decompositions \citep{Kim14} and the total stellar mass of NGC~3351 \citep{Gadotti19}, the bar stellar mass is approximately $2.8\times10^{9}$ M$_{\odot}$.  Part way along the bar ($D \sim 1$ kpc) there is a well defined transverse shell of dust and molecular gas which is orthogonal to the bar feeding lanes, and concentric to the nuclear ring.  This feature has an arclength of approximately 1 kpc and in contrast to the bar, the molecular gas observations (see Section 2.2 below) suggest a gas mass of $\sim 6\pm 4 \times 10^{5}$ M$_{\odot}$ for the shell.  We now highlight below the observations and data analysis steps which we will use to analyse the genesis of this feature and shed light on the evolutionary mechansisms at play in the inner regions of barred galaxies.

\subsection{MUSE Spectroscopy}
Details on the observations, data reduction and analyses of the TIMER MUSE data can be found in Gadotti et al. (2018); here we provide a summary of the most relevant aspects. The observations with MUSE were performed during ESO Period 97, using the normal instrument setup, which yields a spectral resolution of 2.65\AA\ and a spectral coverage from 4750\AA\ to 9350\AA. The image quality during the observations of NGC 3351 was $\sim 0.6''$. We targeted the central $1\times1$ arcmin$^2$ of the galaxy and used dedicated empty sky exposures to remove the contribution from the background.

We used the MUSE pipeline v1.6 \citep{Weilbacher12} and employed the standard calibration plan to reduce the data. In the first step of the data reduction, each science frame is calibrated separately to take out instrumental effects (bias, flat-field and illumination correction).  The spectra are then extracted and calibrated in the wavelength direction using the standard line spread function mapping. The calibrated science frames are combined into the final data cube after performing flux calibration, background subtraction, astrometric registration, correction for differential atmospheric refraction, and resampling of the data cube based on the drizzle algorithm. For some of the science frames, we used a background frame created from interpolating background frames taken before and after the science exposure. This is done to account for variations in the background and best reproduce the background at the time of the science exposure. In the last step of our data reduction process, after employing the MUSE pipeline, we have also implemented a method to improve the removal of residual background emission from the data cube using principal component analysis.

To obtain information from the emission lines in the MUSE cube, such as flux and kinematics (to derive velocity and velocity dispersion), we used PyParadise, which is an extended Python version of Paradise, described in Walcher et al. (2015). We first derived the contribution of stars to the spectra by using Voronoi binning to increase the signal-to-noise ratio to a target of S/N = 40, and then fitted the resulting spectra with a combination of template stellar spectra taken from the Indo-US spectral library \citep{IndoUS}, interpolating over emission lines and residual sky lines. PyParadise also fits the stellar kinematic parameters, and the resulting best-fit spectra are subtracted from the observed spectra spaxel by spaxel. We then fitted all major emission lines, spaxel by spaxel, with Gaussian functions coupled in line width and radial velocity. By forcing the kinematics to match we ensure that line ratios are computed for physically connected regions.  To estimate the errors on all measured parameters 100 Monte Carlo realizations were run, where in each trial noise returned from the MUSE pipeline was used to inject random noise in the spectra prior to the aforementioned analysis steps. 

\subsection{ALMA \& HERACLES Molecular Gas Observations}
To complement our MUSE observations, we used archival data of molecular gas emission in NGC\,3351. In order to obtain gas masses and global kinematic parameters, we used the public CO(2-1) maps from HERACLES\footnote{http://www.cv.nrao.edu/$\sim$aleroy/HERACLES} \citep{Leroy09}. These observations were taken with the IRAM 30m telescope, and recover all the flux at an angular resolution of 13'' (640\,pc at the assumed distance of 10.1\,Mpc; \citealt{Moustakas04}). Given the near-solar metallicity of the galaxy (12+log(O/H) = 8.9; \citealt{Moustakas10}), we use the standard Milky Way conversion factor $\alpha_\mathrm{CO} = 4.4\,M_\odot$\,(K\,km\,s$^{-1}$\,pc$^2)^{-1}$ and assume a fixed line ratio \mbox{CO(2-1)}/\mbox{CO(1-0)}$=0.8$ \citep{Leroy09}.
%  2.6kms velocity resolution

To pinpoint the detailed distribution and kinematics of molecular gas in the vicinity of the nuclear ring, we utilized archival ALMA Early Science Cycle 2 observations from project 2013.1.00885.S (PI: Karin Sandstrom). We downloaded the calibrated and imaged datasets from the ALMA archive\footnote{http://almascience.eso.org/aq}, corrected for the response of the primary beam, and extracted the channels corresponding to the line \mbox{$^{12}$CO(1-0)}, with a rest frame frequency of 115.27\,GHz. These ALMA Band 3 observations reach a typical rms sensitivity of $\sim 6$\,mK per 1.92\,MHz-wide channel (a velocity resolution of 5\,km s$^{-1}$), which results in a S/N$\sim 20-40$ in the star-forming ring, and S/N$\sim 5-10$ in the transverse shell. The synthesized beam is $2.0 \times 1.2"$ ($100 \times 60$ pc; PA$=27^\circ$).

We converted the ALMA cube to brightness temperature and constructed moment maps for the \mbox{$^{12}$CO(1-0)} line: moment 0 corresponds to the integrated intensity, moment 1 is the intensity-weighted velocity field, and moment 2 represents the intensity-weighted velocity dispersion. We integrated emission for channels in the window 114.902-115.031\,GHz, using line-free channels to estimate the noise (remaining channels in the frequency range 114.878-115.068\,GHz). We use a sigma-clipping method for moments 0 and 1, to avoid being biased by noise; our preferred maps use a threshold of $4\sigma$, where $\sigma$ is the rms noise in line-free channels and is calculated on a pixel-by-pixel basis. This means that, for each spatial pixel, we mask any spectral channels with values below four times the rms obtained for that specific position; this is equivalent to running the task \texttt{immoments}\footnote{\url{https://casa.nrao.edu/docs/casaref/image.moments.html}} within CASA \citep{CASA} with the parameter \texttt{includepix} set to a minimum value of 4~rms, where rms is the 2D map corresponding to the standard deviation on line-free channels. It is important to impose a threshold that depends on the position in the map, as noise increases towards outer regions as a result of the primary-beam correction.

We have experimented with different thresholds and different integration windows, confirming that the maps are sufficiently robust for our purposes in the areas of interest.
For the moment 2 map we use the ``window method'' \citep{Bosma81}, which captures more low-level emission, providing a better estimate of the real velocity dispersion. We first impose a $4\sigma$ threshold to identify pixels with significant emission, but then consider all emission in those significant pixels, only limited by a window in frequency. We ignore any spatial pixels below a threshold of $4 \sigma$.  We consider the remaining spatial pixels as significant, and define a range of channels (the frequency window) where emission is considered to compute the second-order moment value.  This window is iteratively expanded (started from the channel of peak emission) until the average continuum emission outside the the window converges according to the criterion from \cite{Bosma81}. The maps agree well with \mbox{$^{13}$CO(1-0)} observations from the ALMA Early Science Cycle 2 project 2013.1.00634.S (PI: Adam Leroy).

\subsection{Ancillary Imaging Data}
Archival HST WFC3 images (Program ID 13364; PI Calzetti) in the F336W and F814W filters were obtained from the MAST archive.  The original observations were taken on 23/04/2014 and subsequently processed with AstroDrizzle. The F814W and F336W images have exposure times of 908 and 1062 seconds, respectively.  A three-colour HST composite image from the LEGUS  survey \citep{LEGUS} is shown in Fig. \ref{fig:hst3col}, with the ALMA CO moment 0 map overlaid.

% SWIFT UVOT Level 3 imaging products (Observation ID:36570001) were obtained from the HEASARC archive.  These UV images were taken 16/01/2008, and have an exposure time of 9174 seconds. 
\subsection{Kinematic Analysis}\label{sect:kin}
To constrain the large-scale CO rotation curve we ran the task \texttt{ROTCUR} within \textit{GIPSY} \citep{Gipsy} on the moment-1 map obtained with HERACLES, which has a much larger field of view than the ALMA observations. Given the relatively low spatial resolution of HERACLES, we assume that the moment-1 map predominantly traces circular rotation and is not too affected by streaming motions. We performed an independent check of the rotation curve by assuming different sets of input parameters for the systemic velocity ($V_\mathrm{sys}$), inclination ($i$), position angle (PA), and the coordinates of the centre ($x_0$, $y_0$). We allowed $V_{sys}$ to vary from 760 to 790 km s$^{-1}$ in steps of 2 km s$^{-1}$. The PA was allowed to vary from 185 to 195 in steps of 1 degree, and a range of $30-60$ degrees was explored for the inclination, in steps of 5 degrees. We initially fixed the kinematic centre visually ($x_{0}=110, y_{0}=110$, corresponding to RA=10:43:57.793, Dec=+11:42:14.20) but we also confirmed that shifting the centre by 2 pixels or more (corresponding to 4") in any directions leads to clearly poorer residuals - specifically when inspecting the residuals of the model velocity field and the observed line of sight velocity maps (($V_{mod} - V_{LOS})^2)$.  We deprojected, derived the corresponding $V(R)$, and compared the final residuals by eye. Both \texttt{ROTCUR} and these tests suggest that the PA is well constrained by PA$\sim (192 \pm 2)^\circ$, $V_\mathrm{sys} \sim (776 \pm 4)$\,km\,s$^{-1}$, and the rotation curve is relatively insensitive to inclination changes ($i \sim 40-50^\circ$).  This agrees with estimates from \cite{Buta88,Tamburro08} and \cite{Mazzalay14} using a variety of tracers.

\begin{figure*}
\begin{center}
\includegraphics[width=0.97\textwidth]{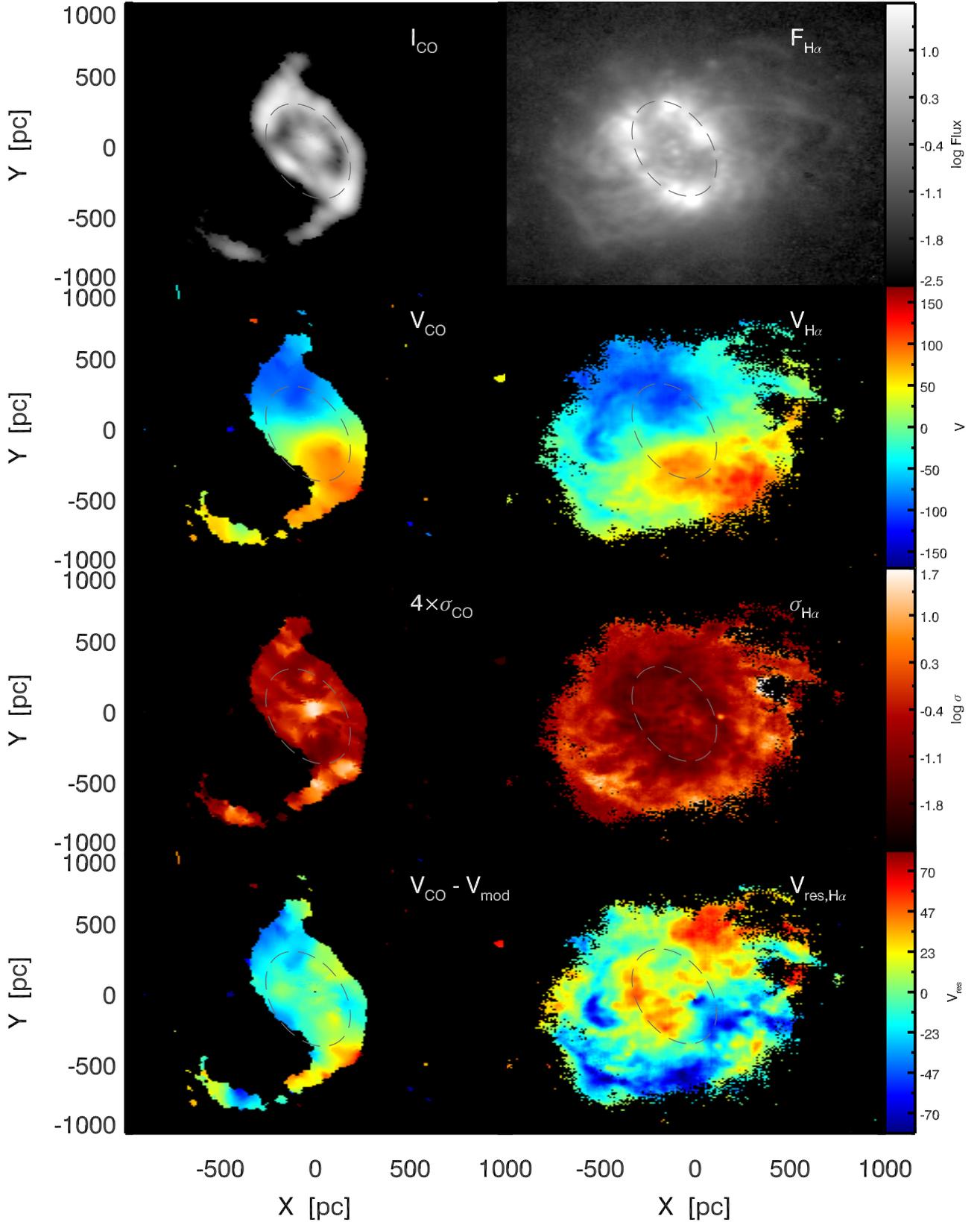}
\caption{\emph{Left column: Molecular gas} From top to bottom are the molecular gas intensity (in K km s$^{-1}$), rotation velocity, velocity dispersion and molecular gas residual velocity ($V_{CO} - V_{circ}$). \emph{Right panel: Ionised gas} H$\alpha$  flux (in erg s$^{-1}$), velocity (in km s$^{-1}$), velocity dispersion and ionised gas residual velocity ($V_{H\alpha} - V_{circ}$; see Eq. 1). An ellipse is reproduced in each panel for reference. Symmetric radial outflows up to $\sim 70$ km s$^{-1}$ are clearly evident in the bottom right panel.}
\label{fig:hamom}
\end{center}
\end{figure*}

The left columns of Fig. \ref{fig:hamom} show the ALMA CO maps of moments 0 (flux), 1 (mean velocity) and 2 (velocity dispersion).  The right column of Fig. \ref{fig:hamom} shows the corresponding H$\alpha$ moment maps obtained from the MUSE data cube. There is filamentary H$\alpha$ emission emanating from the SF ring suggestive of non-gravitational motions.  While the velocity dispersion can reach as high as $\sim$150\,km\,s$^{-1}$, the mean velocity map shows a clear rotation signature.  To discern the amount of rotation and radial/residual gas motions, we apply a harmonic decomposition to the mean velocity field (e.g., \citealt{vdvenfathi10}). We model the velocity field as:
\begin{equation}
\begin{aligned}
&V_\mathrm{mod} = V_\mathrm{sys} + c_{1}\cos(\phi) + s_{1} \sin(\phi) + c_{2}\cos(2\phi) \\
&+ s_{2}\sin(2\phi)+ c_{3}\cos(3\phi) + s_{3}\sin(3\phi),
\end{aligned}
\end{equation}
where the obtained coefficient $c_{1}/\sin(i)$ is the H$\alpha$ rotation velocity and $s_{1}/\sin(i)$ gives the radial flow. Here $\phi$ and $i$ represent the azimuthal angle (from major axis) and inclination respectively. The parameters are obtained by first convolving the model velocity field to the spatial resolution of the observation and then fitting it to the observed velocity field using Levenberg-Marquardt least-squares minimization.
By subtracting the H$\alpha$ model velocity field  (determined from the $c1\cos(\phi)$ term of the harmonic decomposition) from the observed mean velocity map, we obtain a residual H$\alpha$ velocity field. This residual velocity map quantifies any non-circular or deprojected radial velocities in the ionised gas, which may arise due to feedback driven outflows or large scale inflows - in this case $\sim 70$ km s$^{-1}$, planar radial outflows from the ring, which will be discussed in Sect. 3.  We find that the radial component of the velocity field is stably recovered to within $\pm 2$ km s$^{-1}$ when we conduct 200 Monte Carlo perturbations of the observed velocity field with noise.

To further understand whether the high velocity dispersion of H$\alpha$ is dominated by \textit{turbulent} or \textit{gravitational} sources, we leverage our multiple kinematic tracers to perform a decomposition of the H$\alpha$ velocity dispersion ( c.f., \citealt{Weijmans08}). The purely gravitational component of the velocity dispersion can be extracted by comparing the H$\alpha$ rotation velocity to the true circular velocity (as traced by the CO gas or derived from stellar dynamical models).  In the latter case we have derived the circular velocity from the stellar kinematics using Jeans Axisymmetric Models (JAM) \citep{Cap08}.  The stellar dynamical model for NGC 3351 was constructed identical to the descriptions in \cite{Leung18}, and we refer readers to that study for a detailed description.  As shown in \cite{Leung18}, the circular velocities derived from JAM agree well with that derived using CO. This is also the case here, as seen in Figure \ref{fig:dis_decomp} where the red circular velocity curve derived using JAM is in good agreement with the green circular velocity curve derived from CO.. 

The difference between the circular velocity and H$\alpha$ rotation curve is accounted for solely by the gravitational component of the ionised gas velocity dispersion, such that:
\begin {equation}
\begin{aligned}
&V_\mathrm{c}^{2}(R)=\overline{V_\mathrm{\phi,H\alpha}^{2}}+\sigma_\mathrm{grav,R}^{2}\left[\frac{\partial \mathrm{ln} (\nu_\mathrm{H\alpha}\sigma_\mathrm{grav,R}^{2})^{-1}}{\partial \mathrm{ln} R}+\left(\frac{\sigma_\mathrm{grav,\phi}^{2}}{\sigma_\mathrm{grav,R}^{2}}-1\right) \right.\\
&\left. -\frac{R}{\sigma_\mathrm{grav,R}^{2}}\frac{\partial \overline{V_{R}V_\mathrm{z}}}{\partial z}\right]. \\
\end{aligned}
\label{eq_adc}
\end{equation}
Here $V_\mathrm{c}$ denotes the circular velocities for which we have an independent handle from both the CO gas and the stellar kinematics, $V_\mathrm{\phi,H\alpha}$ denotes the H$\alpha$ rotation velocities which we obtain from harmonic decomposition and $\nu_\mathrm{H\alpha}$ is the radial surface density profile of the H$\alpha$ flux distribution. $\nu_\mathrm{H\alpha}$ is well fit by an exponential disk of amplitude $F_{D}$ and scale-length $R_{D}$ superimposed on a Gaussian distribution of amplitude, location and width of $F_{G},R_{G}$ and $\sigma_{G}$ which represents the ring:
\begin{equation}
R_{H\alpha} = F_{D}{\rm exp}(-R/R_{D}) + F_{G}{\rm exp}(-(R-R_{G,2})^{2}/(2\sigma_{G}))
\end{equation}
We find the profile is well described by the parameters ($F_{D},F_{G},R_{D},R_{G},\sigma_{G}$ = 1.99,4.33,1.63,6.2,12.81), where the radial terms are in arcseconds. $V_\mathrm{\phi,H\alpha}$ is parameterised using harmonic expansion of order 2 as described earlier in this section. We also made the assumption that the velocity ellipsoid is aligned with the cylindrical coordinate system. 
With these three ingredients we can solve for $\sigma_\mathrm{grav}$, the gravitational component of the ionised gas velocity dispersion. The turbulent component of the gas dispersion can then be found from $\sigma_\mathrm{turb}^{2}=\sigma_\mathrm{tot}^{2}-\sigma_\mathrm{grav}^{2}-\sigma_\mathrm{therm}^{2}$ (with $\sigma_\mathrm{therm}\sim$10\,km\,s$^{-1}$, assuming a temperature of $10^{4}$\,K).  The results from this velocity dispersion decomposition is most sensitive to the choice of velocity anisotropy than the functional form of the flux profile (which is anyway extremely well behaved).  As such we have marginalised over possible values of the velocity ellipsoid in order to account for the uncertainty in this unknown. The results of this decomposition and the radial profile of the relative contribution of turbulent component to the velocity dispersion profile will be discussed in detail Sect 3.1.3.

% \begin{figure}
% \begin{center}
% \includegraphics[width=0.48\textwidth]{halphavobs.pdf}
% \caption{H$\alpha$ first moment map from the MUSE observations of NGC 3351.}
% \label{fig:f2}
% \end{center}
% \end{figure}

% \begin{table}
% 	\centering
% 	\caption{This is an example table. Captions appear above each table.
% 	Remember to define the quantities, symbols and units used.}
% 	\label{tab:example_table}
% 	\begin{tabular}{lccr} % four columns, alignment for each
% 		\hline
% 		A & B & C & D\\
% 		\hline
% 		1 & 2 & 3 & 4\\
% 		2 & 4 & 6 & 8\\
% 		3 & 5 & 7 & 9\\
% 		\hline
% 	\end{tabular}
% \end{table}

%***********************************************
\section{Hydrodynamical Modelling of the Gas}
%To model the gas dynamics of NGC~3351, we run a suite of hydrodynamic simulations, which are tailored to account for the underlying gravitational potential of NGC~3351. 

As discussed above, and as shown in a number of observational and theoretical studies (e.g. \citealt{Jorsater84,Teuben86,Sandage88,Athanassoula92,Piner95,Kim12,Sormanietal2015}) in barred galaxies gas accumulates along two well defined `primary dust lanes' (see Figure \ref{fig:hst3col}) which run along the leading edges of the bar, funnelling gas to the nuclear region. These form due to shocks induced by the rotating non-axisymmetric gravitational potential, and the morphology of the dust lanes therefore depends on the strength of the non-axisymmetric component of the potential, as well as on the angular frequency at which the bar rotates \citep{Athanassoula92}.

To model the gas dynamics due \emph{purely} to the underlying gravitational potential of NGC 3351, we run a suite of hydrodynamic simulations, which explicitly use the observed mass distribution of the galaxy. While the `primary' dust lanes observed in the galaxy are very likely due to the gas-dynamical response to the gravitational potential, we want to explore whether the curved dust lane is also due to the gravitational potential, or whether some other processes (such as for example stellar feedback) are necessary to create it.  

%-----------------------------
\subsection{Gravitational potential}
\label{sec:pot}
%-----------------------------
The method we use for calculating the potential due to the stellar component involves a straightforward three-dimensional integration over the density distribution of the target galaxy, which is described in detail in \cite{Fragkoudi15}. We first obtain a stellar mass map for the galaxy from the 3.6$\mu m$ NIR image, which is cleaned from foreground stars and is deprojected to be face-on \citep{Querejeta15}. To obtain the stellar mass surface density we assign a mass-to-light ratio of $M/L = 0.6$ \citep{Meidt14, Benny16, Erroz16, Fragkoudi17} to each pixel in the NIR image, which has been shown to be suitable for a wide range of ages and metallicities of the underlying stellar populations \citep{Meidt14}. We then assign a height function in order to account for the thickness of the disc (assuming a relation between the scaleheight and scalelength of the galaxy according to \citealt{Kregel02}), and integrate over the density distribution to obtain the gravitational potential of the disc. We add a dark matter halo with parameters such that the outer parts of the model rotation curve match the observed rotation curve of the galaxy (see Fig. \ref{fig:dis_decomp} and Section 2.4). Therefore the rotation curve derived for NGC 3351 (in Section 2.4) is used as a constraint in building our gas dynamical models.  As has been shown to be the case for massive disc galaxies (e.g., \citealt{Kranz03,Courteau15,Fragkoudi17}) the galaxy is baryon dominated in the central regions, and thus the dark matter potential contributes little to the total gravitational potential in the regions we are interested in (i.e., within 2 kpc).  This setup allows us to initialize a simulation which is tailored to the observed mass distribution of NGC~3351, thus enabling us to study the dynamics of gas in this galaxy. 

%-----------------------------
\subsection{Gas response simulations}
\label{sec:sims}
%-----------------------------

The gas response simulations are produced with the hydrodynamic Adaptive Mesh Refinement (AMR) grid code RAMSES \citep{Teyssier02}. The simulations are run in two dimensions, since, to zeroth order, gas is confined to a thin plane in disc galaxies. The setup used for these gas-dynamical simulations is described in \cite{Fragkoudi17}.

The initial conditions of the simulations consist of a two-dimensional axisymmetric gaseous disc, with a homogeneous density distribution in hydrostatic equilibrium. The disc is truncated using an exponential tapering function\footnote{This function, which has the form $\mathrm{f_{tape}(r)} = \mathrm{exp}((r-r_1)^2)/(2 d^2)$ smoothly truncates the gaseous disc for $r > r_1$, in order to avoid density discontinuities or numerical issues at the boundary of the simulation box.}. The gas is modelled to be isothermal, with adiabatic index 5/3 -- corresponding to HI gas (neutral atomic hydrogen) and is assigned a sound speed of 10 km s$^{-1}$, corresponding to characteristic temperatures of the interstellar medium.

In order to avoid transient features in the disc, we introduce the non-axisymmetric potential gradually over a finite period of time, -- specifically $\sim$3 bar rotations, in accordance with previous studies which calibrated this empirically \citep{Athanassoula92,PatsisAthanassoula00}.
After the non-axisymmetric potential has been fully introduced, the gas settles in a stable state and we use subsequent snapshots to examine the shocks and gas morphology which are produced purely due to the gravitational potential of the system.

We run a suite of simulations in this setup with different values of the bar pattern speed, to find the best-fitting model, \emph{defined as the model which best reproduces the gas morphology of NGC 3351 in the bar region} (i.e. which has primary bar dust lanes in the correct locations\footnote{For more details on this process see Section 4 of \citealt{Fragkoudi17}}). We first scan the parameter space by running five simulations with $\mathcal{R}$ in the range from 1-2 (which is the range of values expected from theory and observations; e.g. \citealt{Athanassoula92,Corsini2011}), where $\mathcal{R}= R_{\rm CR}/R_{\rm bar}$, $R_{\rm CR}$ is the corotation radius and $R_{\rm bar}$ is the bar length. Bars are considered slowly rotating if $\mathcal{R}>1.4$ and fast rotating if $\mathcal{R}<1.4$ (see \citealt{Debattista2000}).  
We found that a fast bar, i.e. with $\mathcal{R}<1.4$ (which corresponds to values of the bar pattern speed between $\sim$33-38 km\,s$^{-1}$\,kpc$^{-1}$) gives a better fit to the gas morphology. We then explore this region of the parameter space, by running five simulations with pattern speeds varying by 1\,km\,s$^{-1}$\,kpc$^{-1}$, to identify more precisely the best fit bar pattern speed. Finally, the best fit model has $\mathcal{R}\sim$1.15 -- which corresponds to a bar pattern speed of $\sim$35 km\,s$^{-1}$\,kpc$^{-1}$. We emphasize that none of the models we explored in this analysis contain a secondary curved dust lane, i.e. even models with much slower or faster bars did not contain such a feature.

\begin{figure}
\begin{center}
\includegraphics[width=0.48\textwidth]{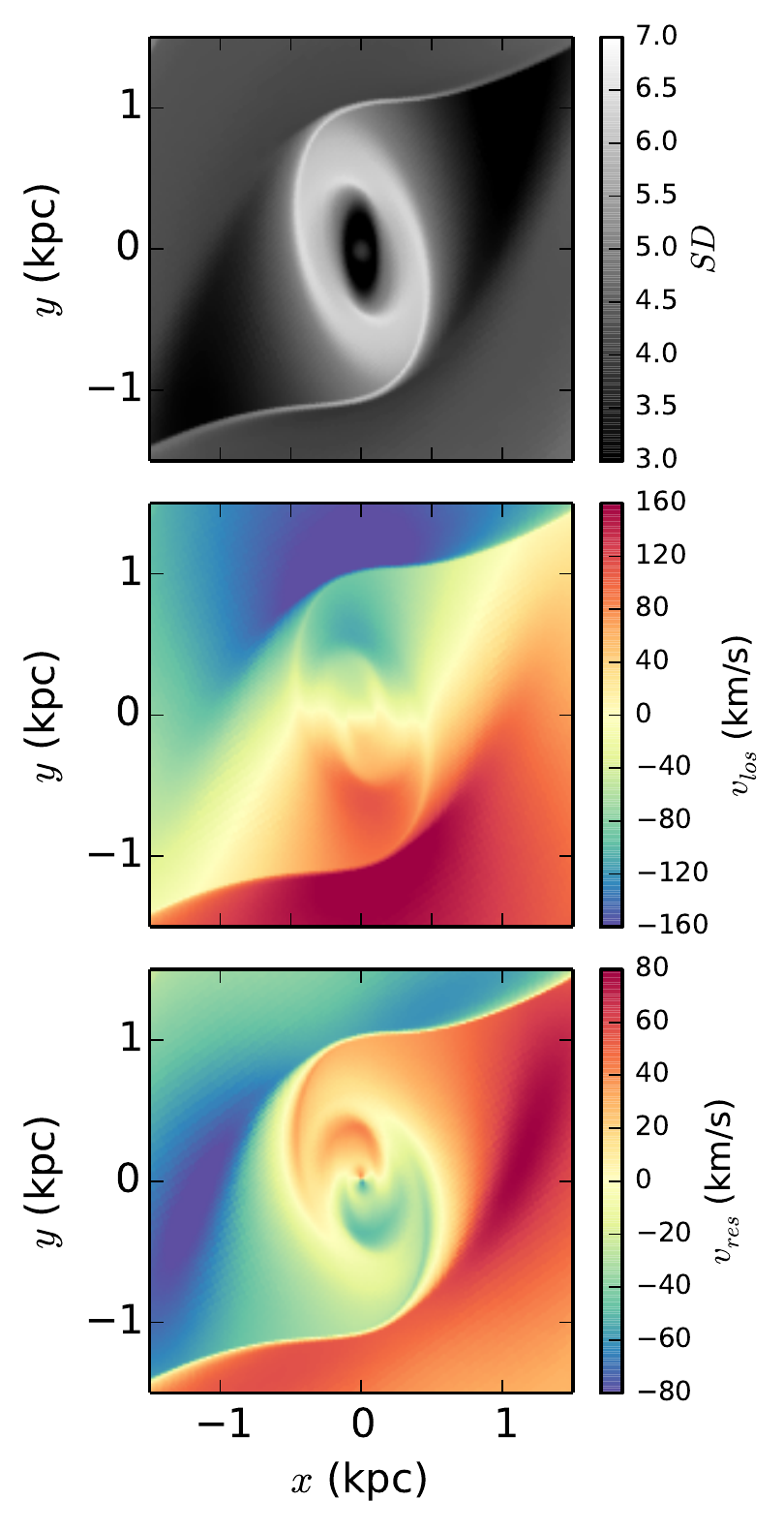}
\caption{Gas surface density for the best fit simulation (in arbitrary units of surface density). There is only one pair of shocks formed at the leading edges of the bar which will produce the linear feeding dust lanes seen in the galaxy -- however there is a clear absence of any analogous transverse dust shells in this purely gravitational simulation.}
\label{fig:sim}
\end{center}
\end{figure}

\subsection{Origin of the Transverse Dust Shell}
\label{sec:origin}

In Fig. \ref{fig:sim} we show the surface density of the gas in the fiducial simulation (top panel), the gas velocity field $v_{los}$ (middle panel), and the residual velocity $v_{res}$ = $v_{los}$ - $v_{circ}$ (bottom panel). In all panels we rotate the model to have the same projection in the plane of the sky as NGC\,3351. 

By examining the morphology of the gas in the simulation (top panel) we see that gas accumulates mainly in the nuclear ring of the galaxy and along the two `primary dust lanes' which are located on the leading edge of the bar. Gas accumulates in shocks along the leading edges of the bar (where dust lanes are formed) due to the torques induced by the underlying non-axisymmetric potential. However, we do not find in any of our models evidence for a secondary curved dust lane. By examining the second panel of Fig \ref{fig:sim}, we see that the overall velocity field of the gas matches quite well the observed velocity field of the CO gas in NGC\,3351 (see Fig. \ref{fig:hamom}).

The fact that the transverse dust lane of NGC\,3351 is not produced in the hydrodynamic simulations suggests that it  is not of a dynamical origin, i.e. it is not caused by the underlying non-axisymmetric potential (as is the case for the primary dust lanes seen on the leading edges of the bar in NGC~3351).  This statement is given special weight in the case of our AMR simulations as they \emph{directly use the derived potential of NGC~3351}. Therefore, additional processes likely related to stellar feedback, which are not captured in the simulations, are necessary to reproduce the observed shocks and gas features. 

As suggested in Sec. 1.1 and discussed in more detail in the next sections, the cold gas currently seen in the transverse dust lane was likely pushed there from the nuclear ring region, by stellar feedback processes. While it is possible that other mechanisms, such as tidal interactions, inflows or the `wiggle' instability, might be responsible for the creation of the curve dust lane, we find these less likely than the scenario we propose, due to the reasons given below: 

As noted in Section 1, NGC 3351 is a relatively isolated galaxy, with its closest neighbour having a projected distance of $\sim$125\,kpc; it is therefore unlikely that the curved dust lane is triggered by a tidal interaction due to a close companion. On the other hand, if there were in-plane inflow through the disc causing this feature, this should be captured by our gas-dynamical model (which models the gas dynamics in the disc due to the gravitational potential). On the other hand, extra-planar flows, such as those caused by fountain flows, can be outwards or inwards moving, however the inwards motion is expected to be small, of the order of $\sim$a hundred parsecs (see e.g. \citealt{Fraternali2008}). Therefore extra-planar gas ejected from star forming regions -- such as the spiral arms -- is unlikely to reach the small radii which we consider here through a fountain flow (see e.g. \citealt{Fraternali2008,Pezzulli2015,DeFelippis2017}). Furthermore, as discussed below in Section 4.2, the residual CO velocities are inconsistent with the CO gas in the transverse dust lane moving inwards. Other effects related to gas instabilities in nuclear rings, such as for example the `wiggle' instabilities described in \citealt{Sormanietal2017}, are also unlikely to produce such large-scale features like the curved dust lane. We therefore conclude that the most likely scenario is one in which feedback from the nuclear ring is responsible for creating the curved dust lane.

Figure \ref{fig:hst3col} shows that, at present day, the cold molecular CO gas and the dust (as traced in optical colours) are co-located at the position of the transverse shell.  This strongly suggests that the initial driving mechanism to move this feature from the nuclear ring outwards, moved the dust and molecular gas together.  In addition the CO gas and dust appear to be evacuated in the area closest to the nuclear ring, suggesting some dynamical movement of these cold dense gas tracers.

Figure \ref{fig:haco2} shows that the subsequent radial expansion of the warm ionised gas has filled the cavity bounded by the outer transverse dust shell.  This low density ionised medium appears nearly completely contained by the extent of the CO/dust shell, as is also suggested by studies which found X-ray emission contained within this cavity \citep{Swartz06}.

% \begin{figure}
% \begin{center}
% \includegraphics[width=0.48\textwidth]{hst3351colo.pdf}
% \caption{Overlay of ALMA CO 1-2 molecular gas moment 0 map (\emph{green contours}) on HST F814W image of the central region of NGC~3351, clearly demonstrating the coupling of the dust and dense, cold gas}
% \label{fig:hstcolo}
% \end{center}
% \end{figure}

%\section{A Testable Feedback Scenario}
\section{Properties of the Ionised and Molecular Gas}
% Overview of scenario: Bar feeds central region with gas through dust lanes. Nuclear ring forms. Feedback pushes out cold gas and creates cavity. Warm gas fills cavity and interacts with cold gas. (Schematic diagram)

%\subsection{Evidence from Data}
Here we seek to use the observations of the molecular and ionised gas kinematics and densities, to understand the dynamical state of the outflow in NGC~3351.  Quantifying the mechanical energy, outflowing velocities and morphological structure of the outflow will allow us to make more rigorous tests for the origin of the outflow in our analytic and numerical models.

\begin{figure*}
\begin{center}
\includegraphics[width=1.02\textwidth]{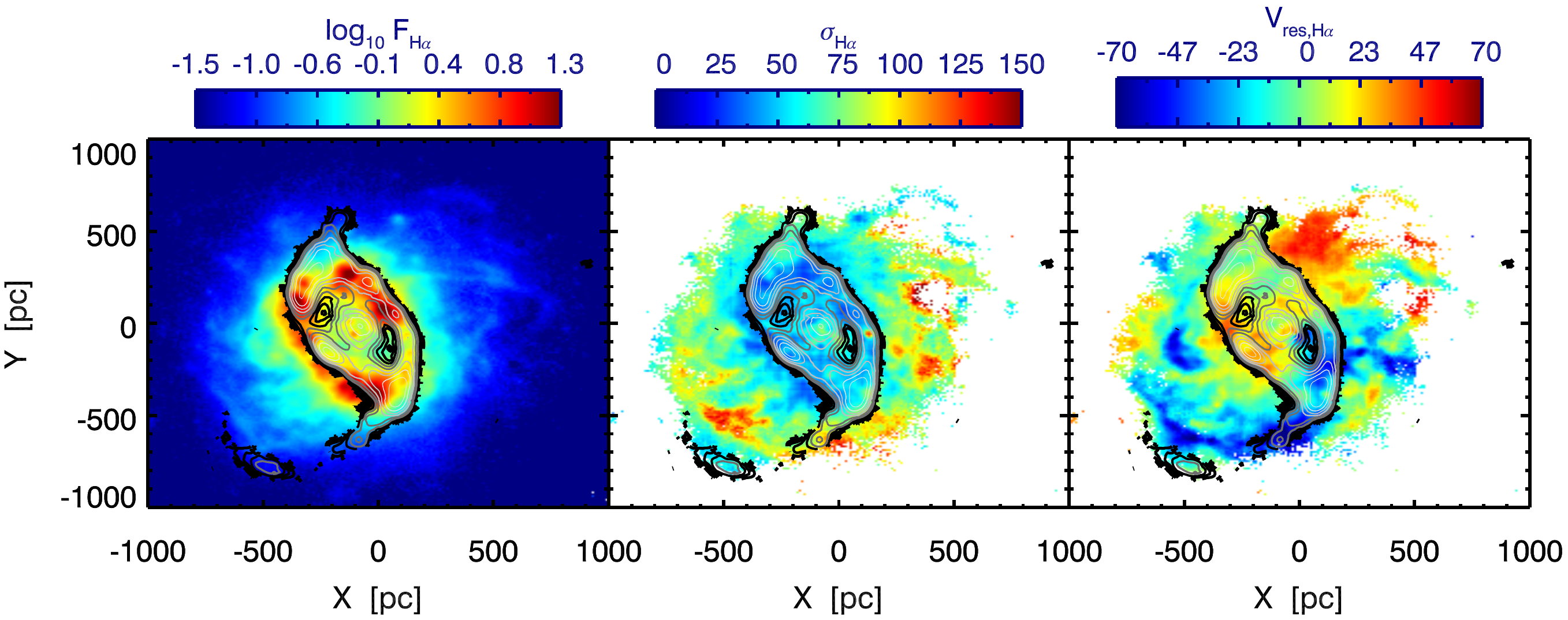}
\caption{H$\alpha$ flux (\emph{left}), velocity dispersion (\emph{middle}) and residual velocity (\emph{right}) fields of NGC 3351 are shown in colour scale.  Overlaid as contours is the ALMA CO(1-0) moment 0 map, which clearly encloses the ionised outflows to the south west of the nuclear ring.}
\label{fig:haco2}
\end{center}
\end{figure*}

%\subsubsection{Dynamical properties of the ionised and molecular gas}
%Halpha velocity and harmonic decomposition and residuals (everything about first moment).\\

From the residual velocity map in Fig. \ref{fig:hamom} and \ref{fig:haco2} we can see a strong bisymmetric radial flow of the ionised gas up to $\sim$70\,km\,s$^{-1}$. As indicated by the ellipses in Fig. \ref{fig:hamom} (which mark the location of the nuclear ring), this radial flow covers the area from the nuclear SF ring (where the H$\alpha$ peak flux is) and passes through a region of high  velocity dispersion near the inner edge of the transverse CO shell.

\subsection{Emission Line Diagnostics}
The bulk of the underlying ionised gas velocity field in Fig. \ref{fig:hamom} has a rotation axis orthogonal to this radial outflow.  Along the bar (roughly parallel to the minor axis of the ring), the LOS velocity field is close to zero -- which suggests that shocks should occur in the cavity where the outflowing gas collides with this stationary (or even partly orthogonally rotating) ISM.

We compute the star formation rate following the \cite{Kennicutt98} relation ($SFR = L_{H\alpha}~1.26\times10^{41} {\rm erg s}^{-1}$), after correcting the H$\alpha$ emission for extinction using the prescriptions in \cite{Calzetti00} and estimating the E(B-V) from the Balmer decrement.  The electron densities were computed using the extinction corrected fluxes of the [SII] doublet emission lines ( at $\lambda = 6716,6731 \AA$) following Equation 1 and Table 1 of \cite{Kaasinen17}.

\begin{figure*}
\begin{center}
\includegraphics[width=0.94\textwidth]{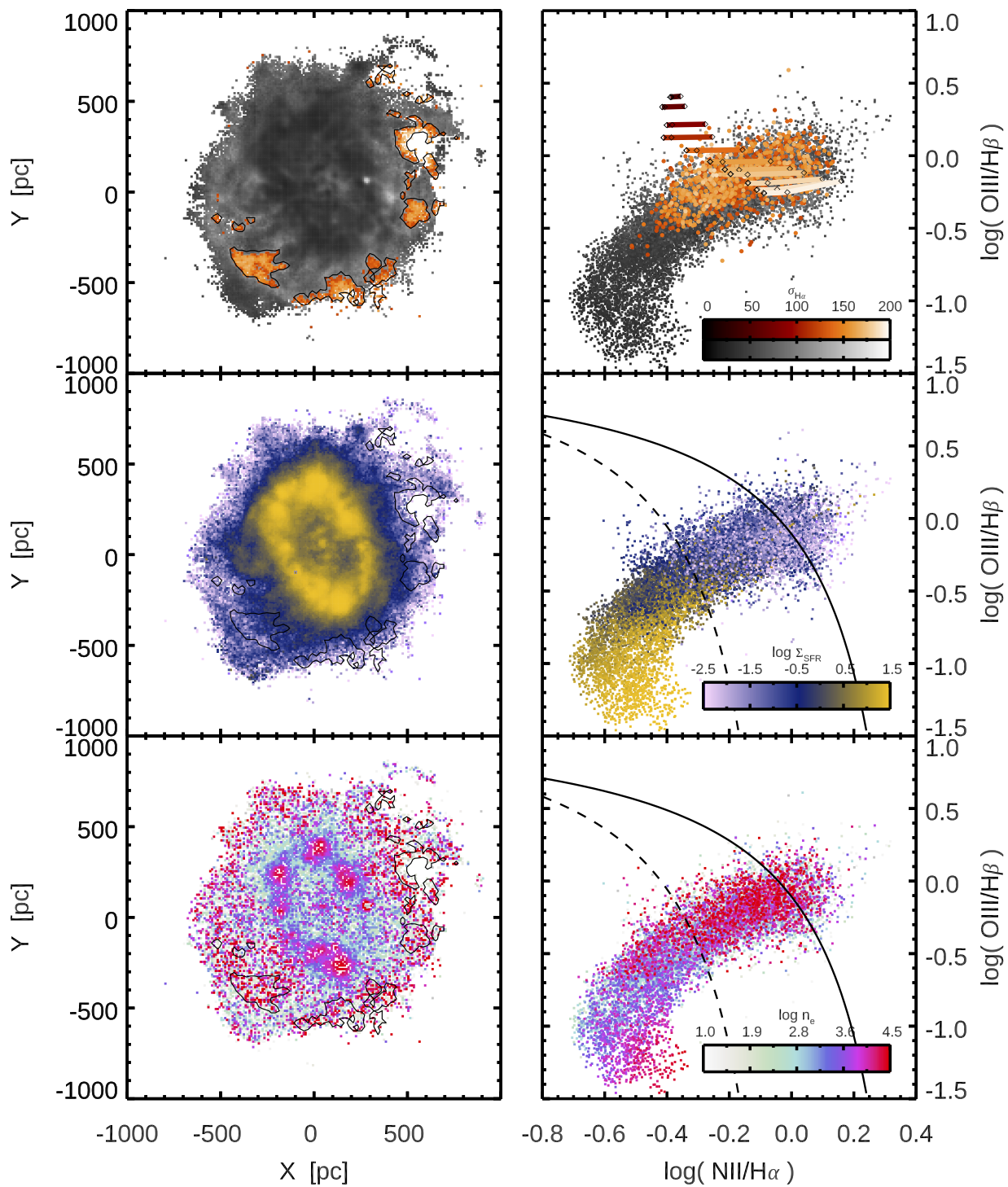}
\caption{Spatial maps (\textit{left}) and BPT diagram (\textit{right}) of the central star forming ring and surrounding area of NGC 3351.  The top panel shows the pixels greyscale coded by $\sigma_{H\alpha}$, with a subset of high dispersion ($\sigma \geq 125$ km s$^{-1}$) pixels shown in the orange colour bar and outlined in black.  These spaxels have emission line ratios consistent with fast shock models of the observed velocity (\textit{coloured tracks on the same velocity scale, spanning from left to right magnetic field strengths from 1 to $100~ \mu G$}).  Middle panel shows the SFR surface density, while bottom panel shows electron density, indicating that the high electron densities in the outer region are due to compressive shocks rather than high SFRs. Solid (dashed) black lines show the upper (lower) theoretical limits expected for SF (AGN) galaxies respectively (e.g.,\citealt{Kewley01,Kauffmann03}).}
\label{fig:bpt}
\end{center}
\end{figure*}

In Fig. \ref{fig:bpt} we show the spaxels in the MUSE field, and their position on the BPT emission line diagnostic diagram (c.f., \citealt{BPT}).  The spaxels outside the nuclear SF ring fall clearly in the shocked region of the BPT diagram (e.g., \citealt{Kewley01,Kauffmann03}).

As a strong consistency check, we can ask if the emission line ratios we observe are consistent with coming from gas with shock velocities close to the ionised gas velocity dispersion we measure.

In the top panel of Fig. \ref{fig:bpt} we show the result of fast-shock models using the MAPPINGS-III code \citep{Allen08}, which predict the emission line ratios for shocked gas of a given density moving at a particular velocity, in a given magnetic field.  The spaxel values for the MUSE emission line ratios are shown in that figure, and the subset of pixels in the high velocity dispersion region of the cavity are colour coded by their velocity dispersion.  The shock models (here run for solar metallicity and densities of $n_{e} = 100 ~{\rm cm}^{-3}$) of different magnetic field strengths (notches along the tracks from $1-100 {\mu}G$) are also colour coded by velocity and show good agreement with respect to the observed velocity and emission line ratios in the cavity.

The BPT diagram emission line diagnostics and shock models indicate that a significant amount of ionizing radiation did not precede expansion of the dusty molecular shell and alter the emission line ratios in the region outside the nuclear ring (e.g., the line ratios are inconsistent with a precursor ionization front).  Precursor shock models for our densities and magnetic fields predict ${\rm log}({\rm O}\textsc{III}/H\beta)$ ratios greater than zero, which are above our measured values (Fig. \ref{fig:bpt}).  This provides important indirect evidence that the bulk of the ionizing radiation in the initial starburst was absorbed by the dusty gas shell and transfered momentum into this structure through direct photon pressure.

The middle panel of Fig. \ref{fig:bpt} shows that the emission line ratios for the region outside the SF ring are indeed in the classical region associated to shocks (c.f., \citealt{Kewley01}). The bottom panels show indications that the cavity contains a few spaxels with electron densities sometimes approaching that of the ring (log~ $n_{e} \sim 4$).  However there is clear stochasticity in the on-sky map which led us to check with both a binned analysis of the MUSE cube and a locally weighted regression smoothing \citep{Cappellari2013} whether these values were representative of the average electron density in the cavity.  In addition we tested whether temperature corrections (using the local velocity dispersion $\sigma_{H\alpha}$ and mean molecular weight $\mu$ to compute a post-shock temperature ($T_{s} = 1.1\mu (\sigma_{H\alpha}/1000)^{2}$ keV) to the electron density would alter the electron density map. All three checks suggested that the electron density between the shell and the ring is higher in regions of large velocity dispersion (even in the $n_{e}$ maps where shock temperature was not used to correct the electron density), and that the typical cavity electron density is closer to $2.0 \lesssim {\rm log}~ n_{e} \lesssim 3.0$.  The stochasticity is likely due to the weak nature of the Sulfur lines used to derive the density and our spectral resolution - despite imposing cuts on the relative flux error for H$\alpha$ ($\leq 0.05$) and the [SII] lines themselves ($\leq 0.5$).   To summarize, the average, error on the mean and standard deviation of the spaxel electron densities in regions of the ring (defined as where the SFR surface density exceeds ${\rm log} \Sigma_{SFR} = 0.5$), the cavity (the high dispersion pixels outlined in red in between the shell and ring in the upper left panel) and the remainder of the area outside of the ring were found to be: {\rm log}$ n_{e,Ring} = 3.07 \pm 0.01 (\sigma = 0.16)$, {\rm log}$ n_{e,Cavity} = 3.05 \pm 0.04 (\sigma = 0.61)$ and {\rm log}$ n_{e,outside} = 2.77 \pm 0.04 (\sigma = 0.60)$. While there is clear uncertainty and difficulty in statistically differentiating the electron densities between the regions - these tests and the spatial correspondance of enhanced electron densities and regions of high velocity dispersion, hint that the values of $n_{e}$ in the cavity region are not due to high SFR surface densities -- instead they may be intrinsic to, and enhanced by, the compressive gas motions which generated the shocks.  This is indirectly given support by the shock models in the top panel, which reproduce the observed line diagnostics when using cavity densities of $n_{e} = 100$, and is also consistent with the presence of X-ray emission in this region \citep{Swartz06}.

% \begin{figure}
% \begin{center}
% \includegraphics[width=0.48\textwidth]{timerp4.pdf}
% \caption{Emission line ratios for all pixels (\textit{grey points}).  Fast shock models are overlaid for different shock velocities (in steps of 25 km s$^{-1}$).  Each notch along a particular velocity model is for a different magnetic field strength.  The pixels in the region of observed high $\sigma_{H\alpha}$ where gas has collided with the CO ring and shocked, are shown as larger circles colour coded by their velocity dispersions.  The location of these relative to the shock model grid, and the agreement between the shock model velocities and observed velocities together provide a strong piece of evidence that the expanding gas running into the stalled colder gas is the source of these shocks. A precursor ionization front is ruled out.}
% \label{fig:mappings}
% \end{center}
% \end{figure}

\subsection{Gas Dynamics}
\begin{figure}
\begin{center}
\includegraphics[width=0.48\textwidth]{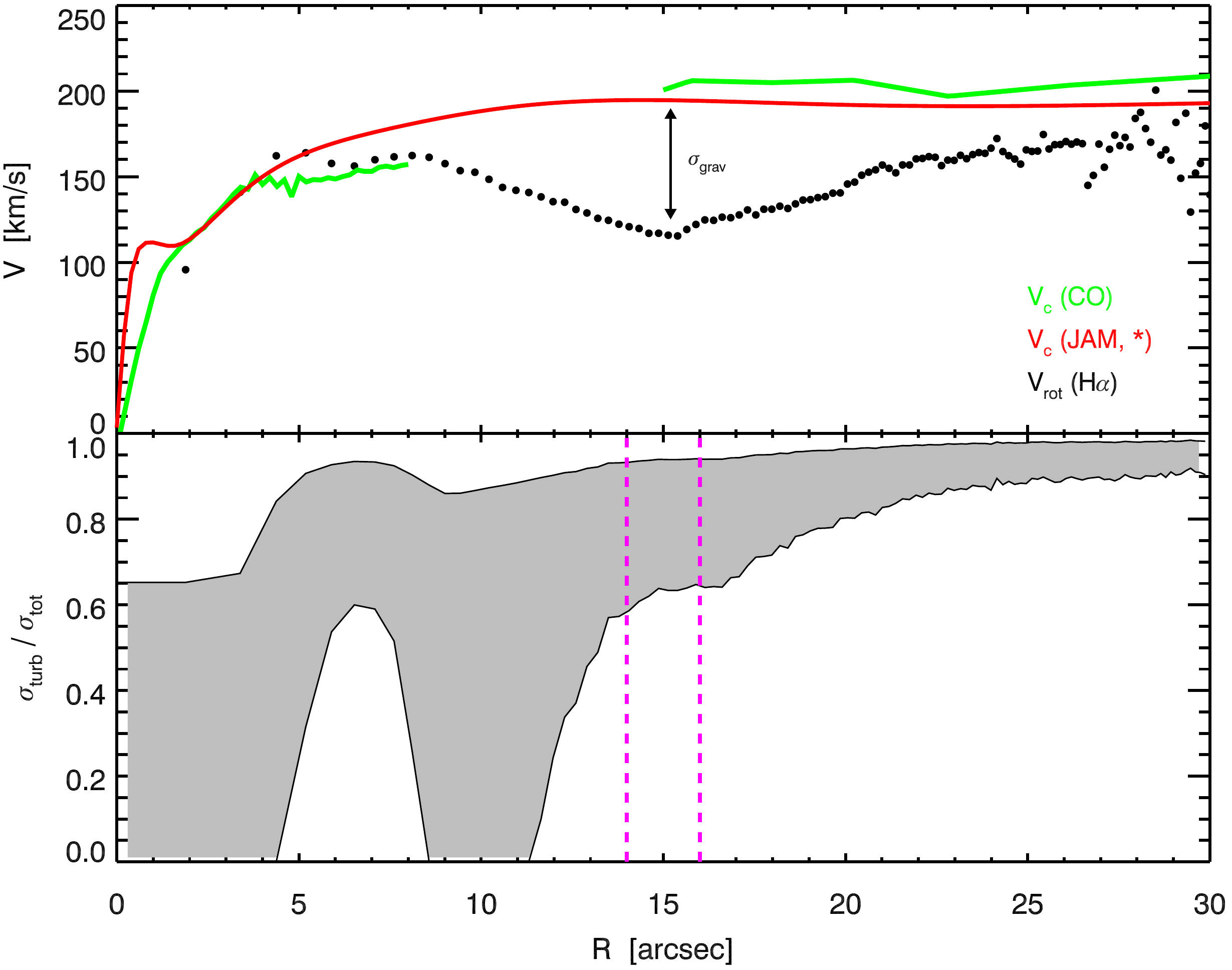}
\caption{\textit{Top panel}: Circular velocities derived from stellar dynamical modeling (red) and CO kinematics (green), allow for a decomposition of H$\alpha$ velocity dispersion into gravitational and turbulent components. Black points indicate the observed H$\alpha$ rotation curve, which we use in solving Eq. \ref{eq_adc} to derive $\sigma_\mathrm{grav,H\alpha}$. \textit{Bottom panel}:  Grey area shows the possible fraction of H$\alpha$ velocity dispersion which is turbulent in nature, with the range due to an uncertainty in velocity anisotropy. Magenta lines indicate the region of the $\sigma_{H\alpha}$ peak.}
\label{fig:dis_decomp}
\end{center}
\end{figure}

The emission line diagnostics suggest a turbulent origin for the large H$\alpha$ velocity dispersion in the cavity.  As an additional check, in Fig. \ref{fig:dis_decomp} we show a decomposition of the H$\alpha$ velocity dispersion into gravitational and non-gravitational (turbulent) components. As described in $\S$ \ref{sect:kin}, this is possible by comparing the H$\alpha$ rotation curve to the ``true'' circular velocity -- as traced independently by the dynamically cold CO rotation curve (and confirmed by dynamical modeling of the stellar kinematics; e.g., \citealt{Leung18}). The ionised gas in Fig. \ref{fig:dis_decomp} is, as expected, rotating slower than this circular velocity.  In the bottom panel we show the derived turbulent fraction of the velocity dispersion, $f_{\rm turb} \equiv \sigma_\mathrm{turb}/\sigma_\mathrm{tot}$, as a function of radius resulting from solving Eq. \ref{eq_adc}. In the region of peak velocity dispersion near the CO and $H\alpha$ interface we see that $f_{\rm turb} \gtrsim 0.6$, suggesting the H$\alpha$ velocity dispersion is dominated by a turbulent/non-gravitational component rather than purely gravitational -- consistent with the above mentioned shock diagnostics. 

% \begin{figure*}
% \begin{center}
% \includegraphics[clip=true, width=0.4\textwidth]{mom2-window-cont.pdf}
% \includegraphics[clip=true, width=0.4\textwidth]{mom1-vrot-cont.pdf}
% \caption{Maps derived from the ALMA CO data. {\it Left}: CO velocity dispersion. A peak in the velocity dispersion of the molecular gas is seen in the south portion of the transverse dust shell. {\it Right}: Residual velocity field derived by subtracting the circular velocity field from the observed velocity field. The transverse dust shell shows an abrupt jump of $V_{res} = -45$ km s$^{-1}$, consistent with the picture in which the expanding warm gas collides and shocks against the dust shell, altering the dynamics of the molecular gas.}
% \label{fig:codyn}
% \end{center}
% \end{figure*}

Shifting focus to the outer transverse dust shell, in the second from bottom left panel of Fig. \ref{fig:hamom}, we show the distribution of the CO velocity dispersion ($\sigma_{CO}$).  This can reach above 30 km s$^{-1}$, but only in three localized regions: the galaxy centre; an expected peak at the southern entry/contact point between the nuclear ring and the bar dust lane; and at the southern portion of the transversal dust lane. The bottom left map in this figure shows the residual velocity field derived by subtracting the circular velocity profile from the observed CO velocity field. The south portion of the transversal dust lane shows projected deviations from circular velocity of approximately $-40$ km s$^{-1}$.%, which could have even larger deprojected values if the residual motion is in the plane of the galaxy.

The detailed kinematics of this transverse CO shell deserve special attention.  The northern half of the feature shows approximately zero residual velocity and very low velocity dispersion, the combination of which suggests that even in the presence of projection effects, this portion of the CO is no longer radially expanding outwards. In the southern portion of the transverse feature we see a discontinuity where the residual velocity field jumps to $V_{res} \sim -40$ km s$^{-1}$, with an associated peak in dispersion.  

Due to inclination effects this decrease in residual velocity could result from a negative azimuthal peculiar velocity or a positive radial component (or a combination of both).
However the northern part of the transverse shell agrees with the velocity expected from the global rotation curve. Given the geometry of the system (the east/left part of the galaxy must be the near side assuming that the main dust lanes appear on the leading side of the bar), if the motion that is causing this negative residual velocity is purely radial, and in the plane of the galaxy, then it can only be moving away from the galactic centre.
It is interesting to note that the largest H$\alpha$ dispersions and radial flows are near this portion of the transverse dust shell.

Taken together in the context of our scenario for the shell evolution, the H$\alpha$ residual velocity and associated dispersion peak may partly arise from the outflowing ionised gas colliding with the slower moving (or stalled) CO shell -- increasing the dispersion in both the CO and H$\alpha$ and inducing small additional radial motions in the southern portion of the transverse CO shell. We discuss further possibilities for the high velocity dispersion, related to the confined outflow of the gas, in Sec. \ref{sec:geom}.

\section{Modelling the Stellar Feedback and Outflow Dynamics}
%\section{Evidence from Modeling the Energetics}
%\subsubsection{Kinetic energy in the expanding gas}
A zeroth order test of whether the stellar feedback in the nuclear ring could be responsible for the outflow comes from considering the kinetic energy of the outflowing gas relative to the energy of the starburst.  Using the velocity field built from the H$\alpha$ emission in the MUSE cube that we derived (Fig. \ref{fig:haco2}), we have determined the speed that the warm gas is expanding radially from the nuclear ring, $v_{rad} \sim 70$ km s$^{-1}$. Using the total H$\alpha$ luminosity, we inferred the total ionized gas mass in the cavity to be approximately $M_{ion} = 1.2\times10^6$\,M$_\odot$ using Equation 2 of \cite{FoersterSchreiber17}. With these quantities we derive an order of magnitude estimate for the kinetic energy imprinted on the warm gas from the nuclear ring, finding $E_{\rm kin} = 1/2 M_{ion}v_{rad}^{2} \simeq 6\times 10^{52}$\,erg. As we shall see below this is approximately two orders of magnitude smaller than the energy provided by the stellar feedback in the nuclear star forming ring (within $R \lesssim 220$ pc), and given that molecular gas mass in the shell is approximately half the gas mass in the cavity, suggests that even considering only energy input from the western side of the ring and a mechanical efficiency of a few percent, stellar feedback could be responsible for the ionised outflow -- and moving the transverse molecular gas feature from the ring, to its present location.

\subsection{Feedback Energetics}
In order to estimate the energy budget from the star-forming nuclear ring of NGC 3351, we combine stellar population synthesis models (\texttt{STARBURST99}; \citealt{Leitherer99}, hereafter SB99) with analytic prescriptions for momentum injection due to various stellar feedback mechanisms.  The baseline SB99 models were run at solar metallicity, with a Kroupa IMF, and a $SFR = 0.6 ~M_{\odot} yr^{-1}$ (which are the average observed metallicity and SFR in the ring derived from our MUSE observations). The models assume the SFR is constant over the relevant dynamical timescales in this system ($\lesssim 100$ Myr).  We use the energy budget and ionizing photon budget from these computations to model four feedback processes: SNe II, stellar winds, direct photon pressure, and photoionisation heating.  Below we describe the baseline model prescriptions, however systematic variations in the input to the SB99 models or in the subgrid feedback prescriptions are discussed in Appendix A.

\subsubsection{SNe II}
The type II SNe rate (SNR) evolves as a function of the starburst age, and the energy due to SNe is output directly by the SB99 models.  The luminosity is computed using the SNR derived from our MUSE data and assuming the energy of the individual SN is $10^{51}$ ergs.  The momentum flux can be expressed generically as \begin{equation}
\dot{p} = \sqrt{\dot{M}\dot{E}}.
\end{equation}
The mass loss for the SNe ($\dot{M}$) is taken to be 10 $M_{\odot}$ per SNe.
Figure \ref{fig:sbsys} shows the momentum flux for the SNe contribution as a function of starburst lifetime from the SB99 models. The significant contribution of these sources becomes apparent after a few Mys, with a negligible metallicity dependence to the timescale and integrated momentum flux.

\subsubsection{Stellar winds}
The energy, mechanical luminosities and momentum flux of the stellar winds are output directly by the SB99 models, and account for radiation driven outflows from stars in post-MS evolutionary phases.  These winds have a strong metallicity dependence as the radiation coupling will be more efficiently absorbed in the envelopes of metal rich stars.  This effect can be seen clearly in Fig. \ref{fig:sbsys} -- however even with the high metallicity regime we see here, the winds from the stars is still sub-dominant.

\subsubsection{Direct photon pressure}
Photons may directly exert momentum against dust grains, with a momentum flux of $\dot{p} = \tau L/c$.  This process may work with different efficiencies in different density regimes, and distances from the source \citep{Reissl18}, however here we try to reproduce the subgrid prescriptions used by cosmological galaxy formation simulations which model this process.  The effect of multiple scatterings of photons within the dust grains can be included (e.g., \citealt{Hopkins11,Rosdahl15}) as well as a metallicity sensitive opacity and density dependence.  A large variety of parameterizations exist, with different physical assumptions for these photon re-scattering and trapping factors.  For our baseline model we mimic the prescriptions used in the FIRE and FIRE-2 simulations \citep{Hopkins14,Hopkins17} which model the physics of absorption and re-emission of a photon in the dusty gas parcel, but do not explicitly have multiple scatterings per photon.  However three other models are illustrated in Appendix A.  The momentum flux is:
\begin{equation}
\dot{p}_{\rm rad} = (1-{\rm exp}(-\tau_{V}))(1 + \Sigma_{\rm gas}\kappa_{\rm IR})\frac{L}{c}.
\end{equation}
The median of the V band extinction values derived from the MUSE cube (Gadotti et al., 2018) were used to compute the optical depth $\tau$ for the luminosity absorption term, while the $\kappa_{\rm IR}$ opacity term uses a linear metallicity scaling of $\kappa_{\rm IR} = Z/Z_{\odot} \times 10~~ {\rm cm}^{2} {\rm g}^{-1}$.  The molecular gas surface density ($\Sigma_{gas}$) was derived from the moment zero CO maps from HERACLES. The luminosity (and energy) of the photoionizing sources are directly output by the SB99 model.

\subsubsection{Photoionisation heating}
The high temperature created by the ionization of the gaseous cavity can exert an effective pressure on the walls of the surrounding region.  Following \cite{Murray10} the effective pressure can be expressed in terms of the ionizing luminosity and temperature as:
\begin{equation}
P_{\rm eff} = 5\times10^{-10}\left(\frac{L}{10^{41} {\rm erg ~s}^{-1}}\right)^{0.5}\left(\frac{5 ~{\rm pc}}{ r_{\rm bub,min}}\right)^{1.5}\frac{T}{8000 {\rm K}}.  %[{\rm dynes~ cm}^{-2}]
\end{equation}
The momentum flux is subsequently expressed as $\dot{p} = 4\pi r_{\rm bub,min}^{2}P_{\rm eff}$.
As we lack auroral emission lines to accurately constrain the temperature we assume the temperature is $10^{6}$ K in the inner regions of the cavity due to the X-ray emission seen \citep{Swartz06}.

\begin{figure}
\begin{center}
\includegraphics[width=0.45\textwidth]{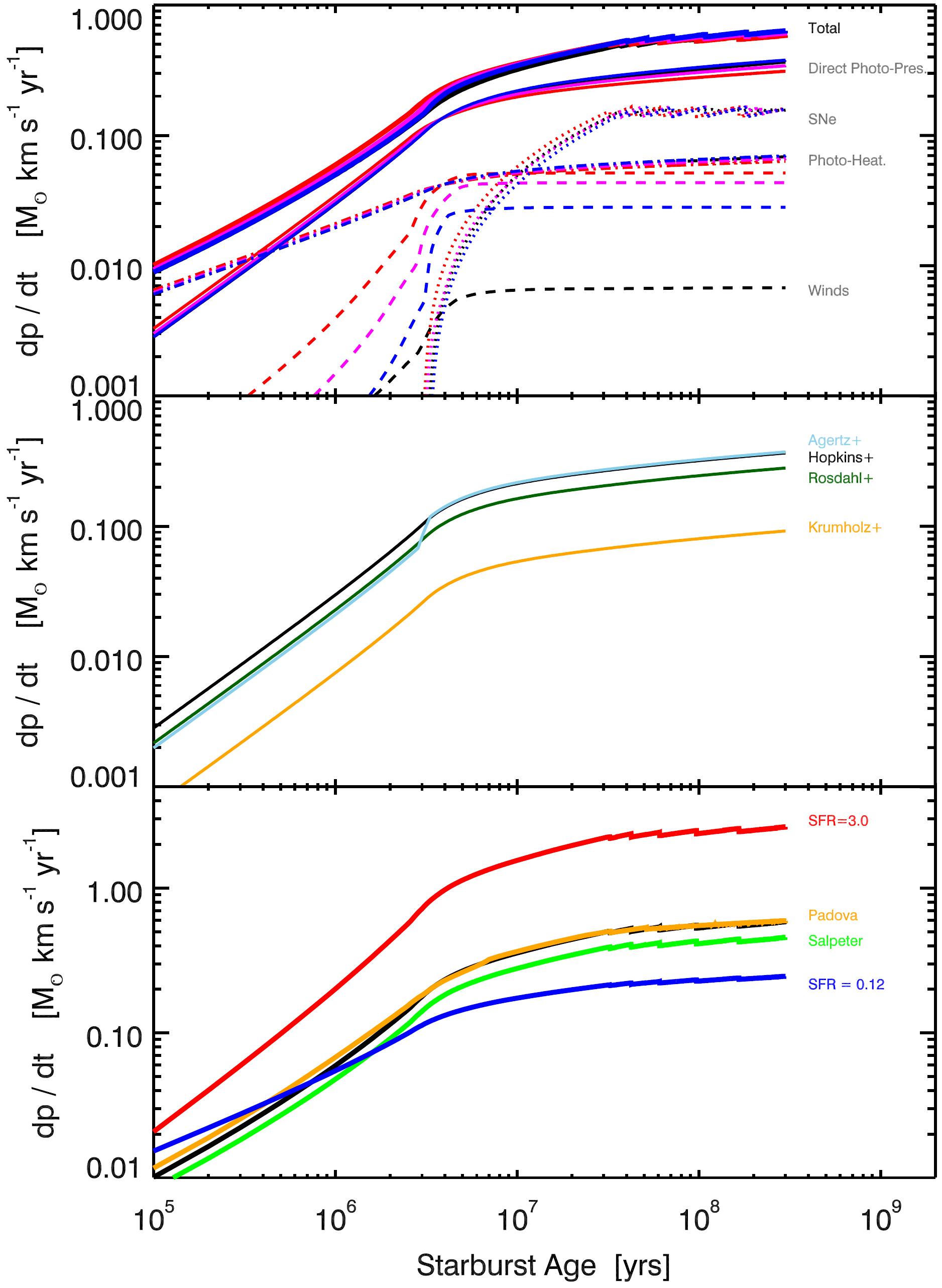}
\caption{Variations in momentum flux due to different physical and sub-grid prescriptions. \emph{Top:} Effect of metallicity changes (\textit{blue}, Z = 0.008; \textit{magenta}, Z=0.020; \textit{red}, Z=0.040) primarily impact the stellar wind contribution. \emph{Middle:} Effect of different sub-grid feedback prescriptions for direct photon pressure coupling and scattering in gas, used by various simulation groups (see Appendix A for details). \emph{Bottom:} Effect of changing SFR in the ring, as well as using a stellar library which includes AGB winds (Padova), or using a Salpeter instead of Kroupa IMF.}
\label{fig:sbsys}
\end{center}
\end{figure}

Figure \ref{fig:sbsys} shows the momentum flux for the four feedback sources.  Under the framework of the parameterizations which we are using (which we have chosen to most closely match cosmological simulation prescriptions) direct momentum injection from the young massive stars ionizing radiation ($\sim \tau L/c$) is the energetically dominant mechanism ($40-60$\%) of the four in this high metallicity regime.  In what follows, given the nearly constant SFH derived for the ring (Appendix B) we utilize momentum flux values in the plateau regions of the curves in, e.g., Figure \ref{fig:sbsys}.  

% \begin{figure}
% \begin{center}
% \includegraphics[width=0.48\textwidth]{timerp2.pdf}
% \caption{Energy, luminosity and momentum of the four feedback processes considered, derived from the \texttt{STARBURST99} models and analytic prescriptions.  Direct photon pressure on the dusty cold ISM plays a significant role in the feedback energy budget.}
% \label{fig:f5}
% \end{center}
% \end{figure}

\subsection{Outflow Expansion Model}
With these feedback prescriptions we can ask what is the typical distance that a parcel of the cold molecular gas can travel as a result of the stellar feedback.  For this exercise the SB99 energetics and feedback momentum fluxes are used in conjunction with analytic models for SN (or any feedback) driven bubble expansion.  We adopt the analytic descriptions in \cite{Kim17} for bubble expansion in a realistic clumpy ISM with continuous energy injection from feedback sources.  These descriptions have been shown to be good reproductions of numerical hydrodynamic results of feedback in galactic disks.  For the pressure driven phase of the bubble evolution (the adiabatic phase is short enough it would occur within a couple of spaxels) the radius evolution can be described as:
\begin{equation}
    R = 34{\rm pc}\left(\frac{\dot{p}}{\rho}\right)^{1/4}\left(\frac{t}{{\rm Myr}}\right)^{1/2}.
\end{equation}
The primary inputs to the models are:  mean density of the ISM ($\rho$), momentum flux ($\dot{p}$) of the explosion/energy source (which comes out of the SB99 and analytic feedback models, and includes dependencies on the SFR, metallicity and dust opacity parameter $\kappa_{IR}$). While a variety of bubble expansion models have been developed in the literature including the seminal work by \cite{ChevClegg85} whose models give comparable results, many of them are tailored towards AGN regimes of high outflow velocity (e.g., \citealt{FaucherGiguere12, Zubovas14}), or have been developed in more detail for specific individual feedback sources (e.g., \citealt{Lagos13,Costa18}).  The adopted models here are flexible enough to be applicable for clumpy multi-phase ISMs for a variety of momentum flux sources, however we anticipate that much more detailed modelling of the dynamics of this system could be done in the future.

For the time being, as our the dynamical arguments and simulations suggest a non-gravitational origin of the transverse shell, we can use these models and now ask \textit{if, and over what timescale, the observed feedback energetics can move the transverse shell of cold gas/dust} from the nuclear ring to its current position.

\begin{figure}
\begin{center}
\includegraphics[width=0.48\textwidth]{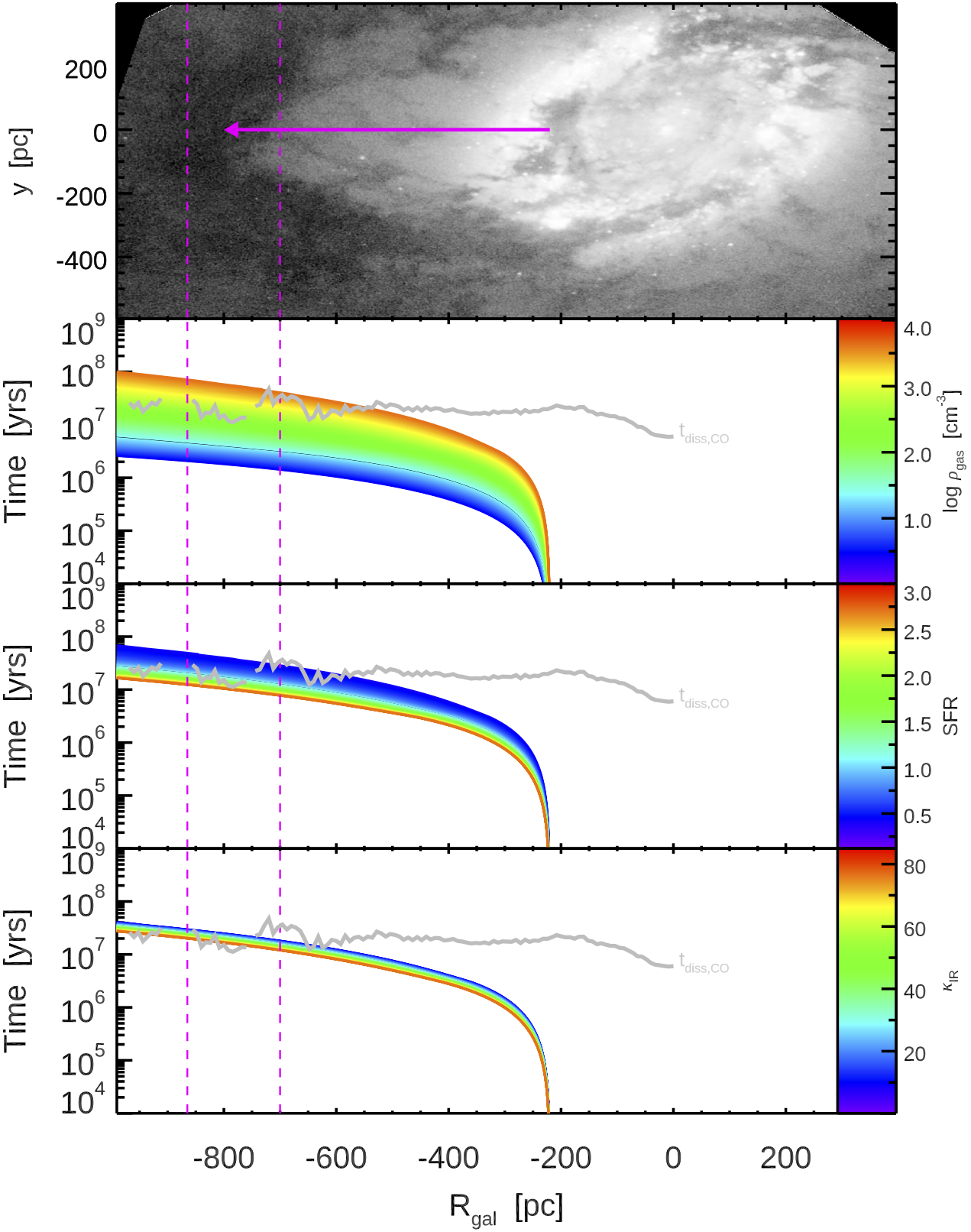}
\caption{Feedback driven bubble evolution models for various initial gas densities (\textit{top panel}), SFRs (\textit{middle panel}) and dust opacity coefficients ($\kappa_{\rm IR}$, \textit{bottom panel}).  The vertical dashed lines show the current location ($D_{CO}$) of the primary dust/gas shell which has been blown out from the SF ring (travelling a distance of $R_{CO}$).  Bottom three panels show the time evolution of the shell for different model parameters (with the other two fixed to their median values). Grey lines show the observed turbulent energy dissipation timescale for the CO gas at that radius -- similar to the likely expansion time for the shell in NGC~3351.}
\label{fig:key}
\end{center}
\end{figure}

Figure \ref{fig:key} shows the time evolution of the dusty molecular gas shell, using the energy derived from the feedback modeling (primarily SNe, photoionisation heating and radiation pressure), in conjunction with simple bubble evolution models. In these illustrative test models the shell typically takes $\sim 10^{7}$ years to expand to its current location, though the three bottom panels in this figure show qualitatively how this changes with differences in the input SFR, surrounding ISM density, or dust opacity factor. The parameterization of the latter mimics multiple scattering events by photons which can boost the momentum deposition by radiation in a dusty ISM. We study the variation of this process, as multi-scattering may be important for our system where the outflow velocities and optical depths are rather low \citep{Costa18}.

Interestingly, the average expansion time for the bubble to travel to its present day location is quite close to the turbulent energy dissipation timescale of the cold molecular gas at this position, as inferred from the CO velocity dispersion (c.f., \citealt{Stilp13}):
\begin{equation}
t_{\rm diss} = 9.8\times10^{6}\left(\frac{\lambda}{100 {\rm pc}}\right)\left(\frac{10 {\rm km s}^{-1}}{\sigma_{CO}}\right)  [yr].
\end{equation}
where $\lambda \sim 100$ pc is a characteristic turbulent driving scale \cite{MaclowMccray88}. Given that a portion of the transverse shell appears radially stalled (Fig. \ref{fig:hamom}), it may suggest that the CO feature is in a semi-equilibrium state, either just on the cusp of breaking out of the disk, or near close to merging with the background turbulent ISM.  If continued momentum injection to the shell front increases the velocity dispersion, it will dissipate on a proportionally quicker timescale. It then may be natural to find a stalling radius where the shell expansion time is closest to $t_{\rm expansion} \simeq t_{\rm diss}$.

To better characterize the expansion time and degeneracies in parameters setting the bubble dynamics, we run an MCMC analysis to ascertain if any qualitative constraints on the original SFR, gas density, metallicity or opacity/re-scattering factor are possible.  The last parameter in particular could be of help for modeling (sub-grid) feedback in modern hydrodynamical galaxy simulations where the softening length or smallest cell size of the simulation is approaching our MUSE spatial resolution ($\sim 10$ pc).

We draw parameters $\theta =$ (SFR, $\rho_{\rm ISM}, Z, \kappa_{\rm IR}$) for each bubble expansion model from a uniform set of priors, which cover the physically expected range of these values in observed or simulated galaxies: $-6 \leq {\rm log} SFR \leq 6$, $-6 \leq {\rm log} \rho \leq 6$, $-3 \leq {\rm log}(Z) \leq 0.7$ and $0.1 \leq \kappa \leq 100$.  The SFR and metallicity set the energy budget output by the SB99 models, which with our feedback parameterizations, describe the effective work done on the shell by SNe, direct photon pressure and subsequent photoionization heating.  The shell is assumed to initially be at the location of the H$\alpha$ SF ring ($R_{\rm initial} = 220 {\rm pc}$, embedded in a density $\rho_{\rm ISM}$.  The range of sampled densities is chosen initially to be quite agnostic and spans values outside those observed for either the ionised or molecular gas in NGC 3351. The available energy from SNe, stellar winds, and direct photon pressure is then used as input along with the gas density to compute the time evolution of the dusty gas shell outwards, using the analytic bubble expansion model (Equation 7).  %To account for continued SFR during the shell's expansion, each model is co-added to an expansion model where the energy is solely produced by photoionisation heating in the cavity, which is incorporated as a contribution during the Sedov-Taylor phase of the expansion before radiative losses take effect (see Appendix B for details).

\begin{figure}
\begin{center}
\includegraphics[width=0.48\textwidth]{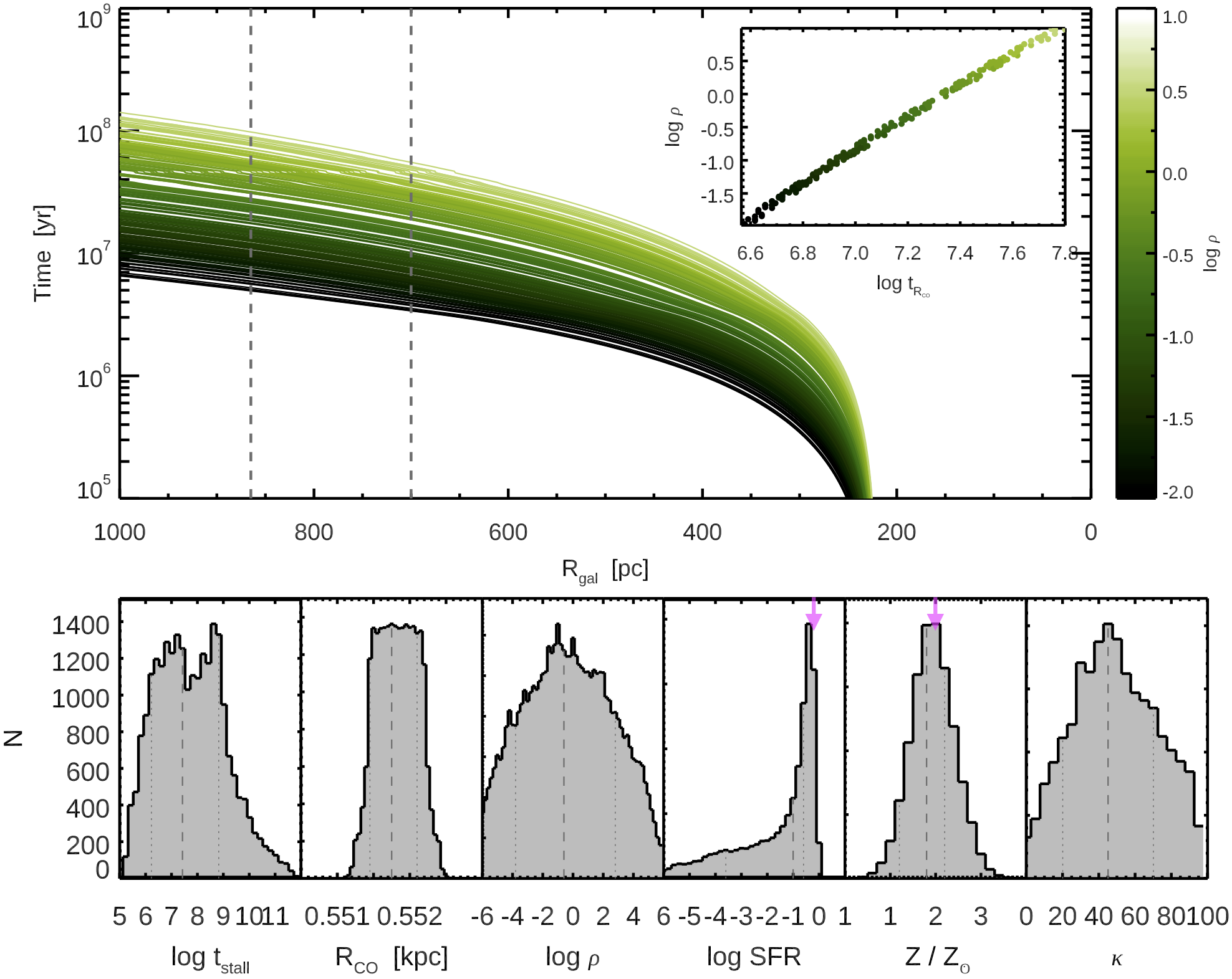}
\caption{Posterior distributions from the MCMC analysis of our feedback driven shell expansion models (\textit{bottom row}).  Median and 1-$\sigma$ confidence intervals are indicated as dashed and dotted lines for the parameters in the model (mean ISM density, SFR, metallicity, and dust opacity coefficient), as well as the recovered shell expansion time at the location of the shell ($R_{CO}$).  Observed SFR and metallicity values in the ring are marked as the magenta arrows.  Top panel shows the distance evolution from the bubble expansion models for trials run for the observed range of densities in NGC 3351, with inset showing the ISM density dependence of the shell expansion time.}
\label{fig:emcee}
\end{center}
\end{figure}

For a given set of model parameters, each evaluation yields a timescale ($t_{\rm stall}$) for the gas shell to travel to the observed present day location ($D_{CO} = 738 \pm 127$ pc), and how close the model makes it to that location ($D_{max}$) (as some never pass that distance).  The likelihood is computed taking into account the difference between the model metallicity, SFR and maximum distance, and those observed quantities ($D_{CO} = 738 \pm 127$ pc, SFR$ = 0.6 \pm 0.3$, $Z/Z_{\odot} = 2 \pm 0.5$). To sample the distributions we use a modified \texttt{IDL} implementation of \texttt{emcee}, originally developed for \texttt{python} by \cite{emcee}, with 300 walkers running 1000 simulations each. The first 30\% of the trials are discarded in order to ensure the results are not dependent on the walker initialization.  The posterior distributions for the four model parameters ($SFR,\rho_{\rm ISM}, Z$, and $\kappa_{\rm IR}$) as well as the time when expansion crosses the observed location of the CO shell are shown in Fig. \ref{fig:emcee}.  Covariance plots between the model parameters are shown in Fig. \ref{fig:covar} and are discussed in Appendix C. %The likelihood for each model was computed by comparing the model bubble's final radius ($R_{\rm final} = d_{\rm stall} + (R_{\rm initial} = 220 {\rm pc}$); where $R_{\rm initial}$ is the current galactocentric radius of the central $H\alpha$ SF ring) --  with the observed present day CO shell galactocentric radius, $R_{CO} = 738 \pm 127$ pc.

\begin{figure*}
\begin{center}
\includegraphics[width=0.98\textwidth]{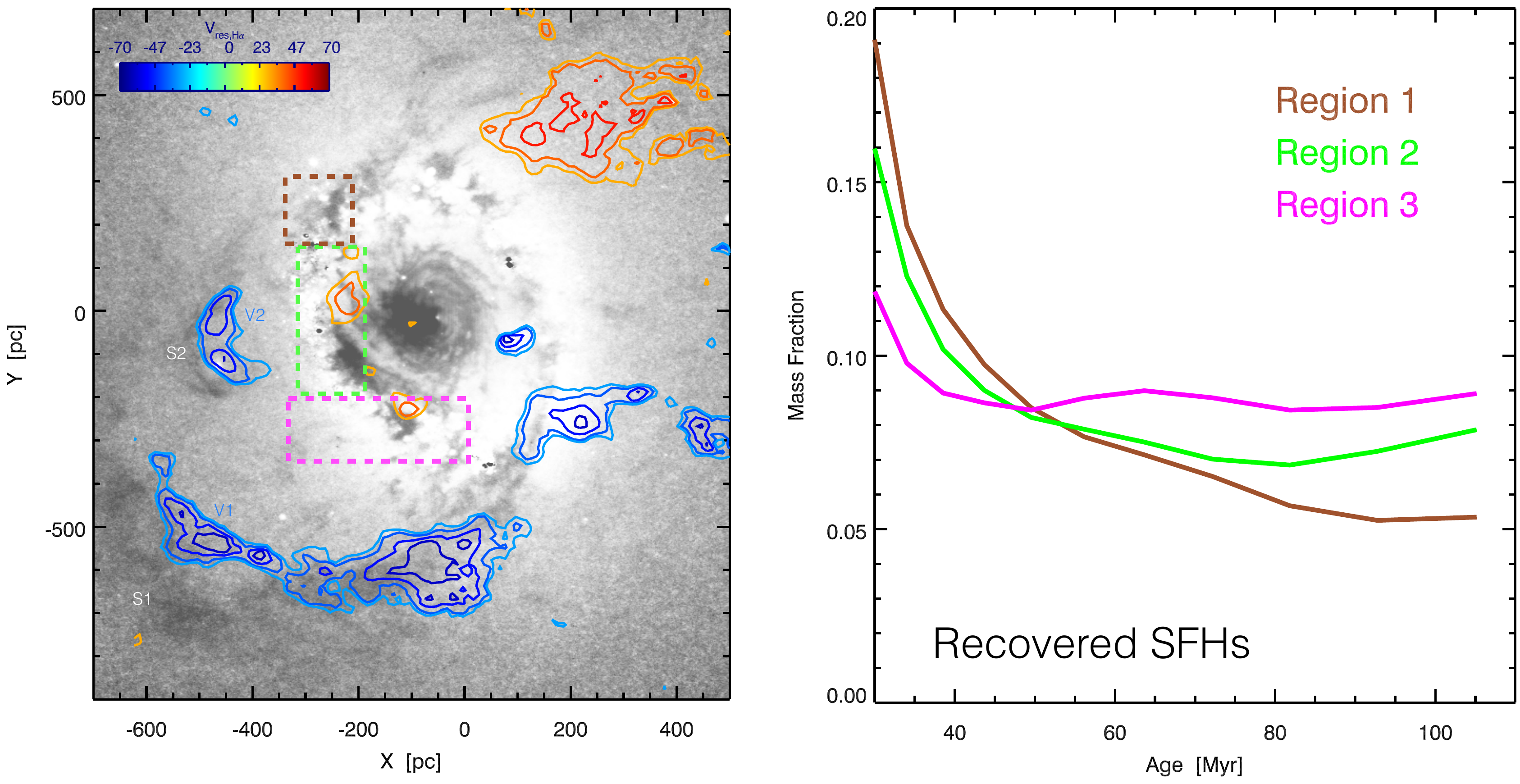}
\caption{HST F555W-F814W greyscale image (\textit{left}) showing the primary dust shell (S1), as well as a smaller secondary dust feature (S2).  The secondary shell has an associated ionised gas outflow kinematic signature ($V_{\rm res,H\alpha}$ shown as coloured contours) and shows a distinct SFH (\textit{right}).}
\label{fig:sfh}
\end{center}
\end{figure*}

The top panel of Figure \ref{fig:emcee} also shows some of the most likely bubble expansion models which reproduce the CO shell location.  Here we have reproduced only models with densities spanned by the ionized and molecular gas observations in the galaxy.  In these models the shell expands to its current location in a typical timescale of log $t_{\rm stall} = 7.2 \pm 0.6$ yr.  Even for the limiting kinetic energy values seen in the hot gas ($\sim 6\times10^{52}$ erg), this implies a total power of up to $4.8\times10^{38}$ erg  s$^{-1}$. The favoured initial density and metallicity are quite close to those derived from our observations, though the SFR is a factor of $1.5-6$ lower than the integrated SFR in the H$\alpha$ ring.  However we measure the present day SFR over the full ring (not the half which impacts the outflow), and our derived SFHs (see Fig \ref{fig:sfh} and Appendix C) suggest the SFR might have been lower in the past.  It therefore appears that a stellar feedback origin for the transverse dust and CO shell is quantitatively supported by these simple models.

\subsection{Initial Mass Loading Factors}
\begin{figure}
\begin{center}
\includegraphics[width=0.48\textwidth]{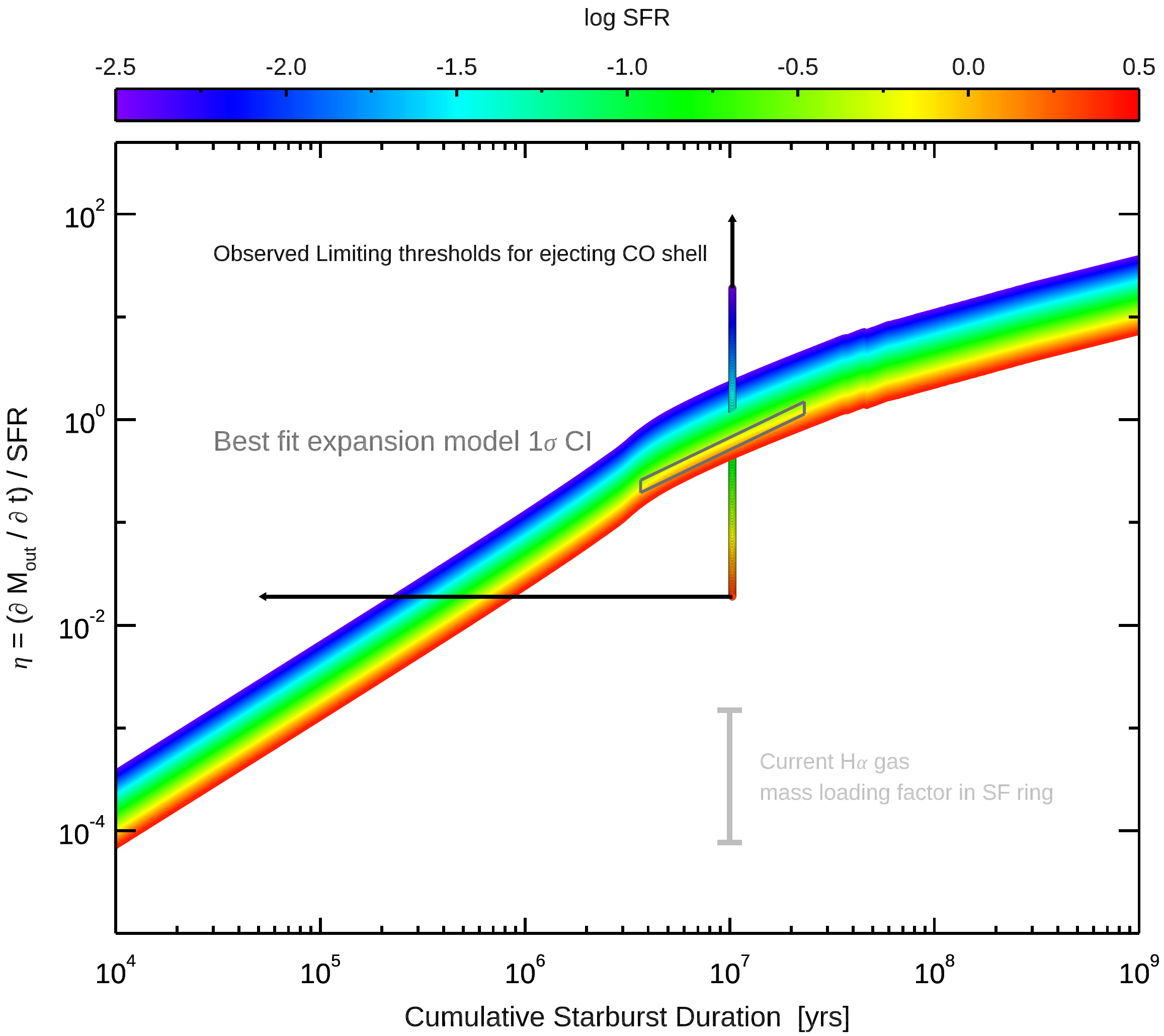}
\caption{Mass-loading factor (mass outflow rate divided by SFR) as a function of time from the combined SB99, feedback and bubble evolution models (coloured lines).  Shown as the grey error bar is the current mass-loading factor for the low density H$\alpha$ gas in the cavity between the CO shell and  nuclear ring.  Black solid line shows the necessary mass loading factor that would have been required to initially eject the cold molecular gas shell from the ring to its observed present day position if the current outflow velocity is used. Dark grey polygon is the $1\sigma$ region from our best fit bubble expansion models.}
\label{fig:massload}
\end{center}
\end{figure}

Simulations of MW mass galaxies can show redshift zero mass loading factors ($\eta = \dot{M}_{\rm out}/SFR$) of below unity, while at high redshift they may approach 10 \citep{Muratov15}.
Our combined feedback and outflow models yield the distance-time evolution of the shell, for a given model's initial density and SFR.  This provides all the necessary ingredients for an estimate of the \textit{initial} mass loading factor, which self-consistently powers the expansion of the shell to the current location. These are shown schematically as the coloured bands in Fig. \ref{fig:massload} for different input SFRs in a model. 
%The mass loading factor implied by our most likely feedback and bubble evolution modeling is shown in Figure \ref{fig:massload} for a range of integrated SFRs.

%****\/\/\textbf{HERE AND RIGHT EQUATIONS AND DESCRIBE FIGURE BETTER}

%From our feedback and outflow models we have evolution of the momentum flux $\dot{p} = \dot{M}_{out}V$ and SFR as a function of time.  Combining this with the bubble expansion models, which give $R_{bub}(t)$ and therefore $V(t)$, we derive the mass loading factor for a given model.  

%More relevant is to consider what $\eta$ in the nuclear ring would have been in order for the molecular gas shell to expand to its current location.  For this \emph{limiting case}, we use the observed molecular gas mass in the shell, and assume outflow velocity seen in the cavity (50 km s$^{-1}$; Figure \ref{fig:sfh}) is constant as the shell expands to the present day location.  For a given SFR, these quantities provide an observational upper limit to the expansion time, and lower limit to the mass loading factor necessary for the shell expansion.  

A conservative estimate of the initial mass loading factor can be taken from the observed molecular gas mass in the shell and a constant observed outflow velocity of $50$ km s$^{-1}$ ($\eta \sim (M \times V/R)/(SFR)$).  This is indicated in Fig. \ref{fig:massload} as black lines. We note that this conservative lower limit assumes a spherical outflowing geometry, while most of the molecular gas is concentrated in a shell.  At the radius of the molecular gas shell, the mass loading factor would have to be multiplied by $\Omega_{sph} / \Omega_{obs} \sim 3.5$ with respect to the spherical analytic models in order to account for this geometrical difference.%\textbf{We note that the bubble expansion models assume a spherical outflowing geometry, while the limits implied by the simple estimate using the observed molecular gas is computed solely for the geometry of the shell.  At the radius of the molecular gas shell, the mass loading computed from the observed shell mass (black lines) factor would need to be corrected by a factor of $\sim 2.7$ with respect to the spherical analytic models in order to account for this geometrical difference.}

From the shell expansion models, we can coarsely estimate the initial mass loading factor at the time the shell was launched by considering the input SFR for the model, the total momentum flux of all feedback sources incident on the shell, and the model derived velocity at each point in time ($\eta \sim (dp/dt)/v(t)/SFR$).  This approximate estimate of the mass loading factor assumes spherical symmetry for the long-term evolution of the shell through the ISM, but does not specify whether the shell was launched as a coherent entity, or swept up along the way - as the model is noted to be well tailored to simulations of a heterogeneous and clumpy ISM.   The shell expansion models which fall within $1-\sigma$ of the observed SFR ($0.6 \pm 0.3$ M$_{\odot}$ yr$^{-1}$) suggest an initial mass loading factor of $0.2 \lesssim \eta \lesssim 1.5$ (\textit{grey parallelogram}), similar to what is seen in simulations of z=0 MW mass galaxies in literature (c.f., \citealt{Schroetter16}).Reassuringly, these models lay within the permitted region on this diagram -- providing a strong consistency check.
 
The initial mass loading factor can be compared to observational estimates of the \emph{current} mass outflow rate from the ring in NGC~3351 following Equation 2 of \cite{FoersterSchreiber17}.  
% The initial mass loading factor can be compared to observational estimates of the \emph{current} mass outflow rate from the ring in NGC~3351 following \cite{Bouche12}, who give, for a transverse sightline:
% \begin{equation}
% \dot{M}_{\rm out} = \frac{\Sigma_{\rm gas}b}{2\theta_{\rm max}}V_{\rm out}2\pi[1 - {\rm cos}\theta_{\rm max}].
% \end{equation}
We compute this present day mass loading factor locally in the ring using the current ionised gas mass, and the maximal observed radial component of the ionised gas velocity field ($V_{\rm rad} \sim 70$ km s$^{-1}$).  This instantaneous mass loading factor for the ionised gas is well below unity ($\eta_{0} \leq 1.3\times10^{-3}$), not surprising given its low density.  

% \begin{table}
% \caption{Feedback+Dynamics Modeling Posteriors}
% \label{table:pars}
% \begin{tabular}{@{}lccc}
% \hline
% %\tablehead{
% Variable & 16th\% & 50th\% & 84\%\\%}
% \hline
% \hline
% $log_{10}~t_{800}$ & 6.67 & 7.17 & 7.67\\
% \hline
% $log_{10} \rho_{ISM}$ & -0.17 & 0.93 & 2.03\\
% \hline
% $log_{10} SFR$ & -1.21 & -0.30 & 0.50\\
% \hline
% $(Z/Z_{\odot})$ & 1.07 & 2.27 & 3.77\\
%  \hline
% $\kappa_{IR}$ & 17 & 44 & 71\\
% \hline
% \end{tabular}
% \end{table}

%\subsection{Evidence from Simulating the Gas Dynamics}
%As described in $\S 2.5$, we run a suite of gas-dynamical response simulations in which we explore the shocks that are induced in the gas solely due to the non-axisymmetric potential of the galaxy. The simulations do not include any star formation or feedback prescriptions, and as such only capture the effects of the underlying gravitational potential on the gas dynamics. 

%Our simulations and others from literature would suggest that the curved dust lane visible in the galaxy is not of a dynamical origin, i.e. it is not caused by the underlying non-axisymmetric potential (as is the case for the main linear dust lanes seen on the leading edges of the bar in NGC~3351 and many barred galaxies).  This statement is given special weight in the case of our AMR simulations as it \emph{directly uses the observed potential of NGC~3351} as an initial condition, and still no such transverse dust lane is visible in the simulations in Figure \ref{fig:sim}. Therefore, additional processes likely related to stellar feedback, which are not captured in the simulations, are necessary to reproduce the observed shocks and gas features.

\section{Survival and Fate of the Cold Gas}
While molecular and neutral gas streamers and shells have been observed in much larger, galactic-scale winds (e.g., \citealt{Walter17}), the genesis of cold molecular gas in any energetic, ionised outflow is still debated.  The cold gas is naively expected to be crushed by shocks or heated by conduction on timescales much less than the outflow travel times.  Here we consider three theoretical models for cold gas survival/formation in outflows in the context of the molecular shell which bounds the cavity of H$\alpha$ emission (S1 in Fig. \ref{fig:sfh}), and which we interpret as having been initially pushed out by stellar feedback.

\subsection{In-situ Condensation}
\cite{Mccourt18} recently posited that a cold component may condense in-situ in the outflow if the warm ionised gas can shatter into droplets, which isobarically cool on short timescales.  The ratio of the cooling time to the crushing time for any cloud is proportional to the Mach number, $t_{\rm cool}/t_{\rm crush} \propto \mathcal{M}$.  Therefore as our outflow is supersonic ($\mathcal{M} \simeq 2$; see \S 6.2), we don't expect this condensation to occur -- and even if it did, the entrainment time (for the condensed gas to move with the hot flow) would be a factor of three longer than the crushing time (see Table 1).  Thus both the survival of the gas and its entrainment are not likely to happen in the necessary timescale to allow for in-situ condensation of gas in this particular outflow.  In addition, the authors note that this process should only help transition ionised gas to a neutral phase, and may not apply to supersonic, self-gravitating molecular gas.  Post-shock cooling scenarios developed for molecular AGN outflows (c.f., \citealt{King15}), require orders of magnitude higher shock velocities and cavity temperatures than seen here. Therefore we consider that in-situ condensation is not a highly probably origin for the molecular gas shell in NGC~3351.

\subsection{Radiative Cooling}
A radiative cooling scenario (e.g., \citealt{Thompson16}) allows for direct cooling of the ionised gas, under certain wind mass loading factors.  For our derived \textit{initial} mass loading factor (see $\S 5.3$) $\eta = 0.85 \pm 0.6$ and given our measured SFR and projected distance of the shell, we can solve for the cooling and advection times of the gas following \cite{Thompson16}:
\begin{eqnarray}
{\rm log}~ t_{\rm cool} &\propto& \eta^{-3.2}(R_{i}/R_{f})^{0.27}R_{i}^{2}/SFR = 7.9_{-0.8}^{+2.0}\\
{\rm log}~ t_{\rm adv} &\propto& \eta^{0.5}R_{f} = 5.8_{-0.4}^{+0.2}
\end{eqnarray}
where $R_{i}$ and $R_{f}$ are the initial and final radii of the shell (taken to be the radius of the nuclear ring, and present day location of the CO shell respectively).  The range of values incorporates the uncertainty on the final shell distance and the initial mass loading factors (Section 5.3).

\cite{Thompson16} then give the critical radius where radiative cooling would take over as:
\begin{equation}
R_{\rm cool} \propto \eta^{-2.92}R_{i}^{1.79}SFR^{-0.789}
\end{equation}
which when using the \textit{maximum} allowable range of $\eta$ from our models in Fig. \ref{fig:massload} gives for our system:
\begin{equation}
0.13 \leq R_{\rm cool} \leq 2322 ~~~{\rm kpc},
\end{equation}
At first glance, the uncertainties in the modeled initial mass loading factor do not constrain this cooling radius.  Similarly, if we ask the inverse question -- what would be the SFR and mass loading factor necessary to have the cooling radius at the observed location of the CO shell, we find:
\begin{eqnarray}
&\eta_{\rm crit}& \propto R_{\rm CO,obs}^{-0.342}R_{i}^{0.613}SFR^{-0.27} \geq 2.97\\
&\log \Sigma_{\rm SFR,crit}& \propto R_{i}^{0.267}R_{\rm CO,obs}^{1.27}\eta^{-3.70} \gtrsim 1.6
\end{eqnarray}
where the latter is in units of M$_{\odot}$ {\rm yr}$^{-1}$ {\rm kpc}$^{-2}$.

\subsubsection{Independent estimates of the initial mass loading factor}
The critical mass loading and SFR surface density reported in Equations 13 and 14 necessary for radiative cooling to dominate, are only marginally outside the range of the values of $\eta$ from our bubble expansion models (e.g., the grey parallelogram in Fig. \ref{fig:massload}), and the SFR densities at peak locations in the ring (Fig. \ref{fig:bpt}).  Lacking more stringent constraints from that dynamical analysis and wishing to better understand whether radiative cooling could play a role in the outflow, we here explore two additional independent constraints on the mass loading factor.  The first is a theoretical estimate for the fraction of escaping mass in the outflow, and the second is a related observational estimate of  the fraction of mass in our outflow which is \textit{above} the escape velocity curve of the galactic potential - and both can be related to the initial mass loading factor.

As outlined in \cite{Thompson16b}, even regions of the ISM that are impacted by sub-Eddington momentum injection can experience escape from the system if they are in under-dense regions of the ISM.  The Eddington ratio is then implicitly related to a critical density in the ISM, such that average densities ($\langle \Sigma \rangle$) below this are susceptible to ejection:
\begin{equation}
\langle \Gamma \rangle \equiv \frac{L}{\langle L_{\rm edd} \rangle} = \frac{L}{4\pi G c \Sigma_{CO}} =\frac{\Sigma_{\rm crit}}{\langle \Sigma \rangle}
\end{equation}
For the ring's H$\alpha$ luminosity of $\sim 6\times 10^{39}$ erg s$^{-1}$, and assuming that the peak gas density seen in the shell ($\Sigma_{CO} \sim 5- 30$ M$_{\odot}$ pc$^{-2}$) was initially co-located in the star forming ring at the time of its launch, we estimate a broad possible range for the ratio $0.005 \leq \Gamma \leq 0.03$.
With these \cite{Thompson16b} provide an estimate of $\varsigma$, the mass fraction of gas that will be ejected from the potential - given our observed Eddington ratio and Mach number for the outflow.   The initial Mach number of the flow is difficult to know, so we consider the range exhibited by the current gas kinematics in the galaxy.  For the ionized component this is $\mathcal{M} \equiv v/c_{s} = 1.9$ if we use the ionized gas outflow velocity ($v = 70$ km s$^{-1}$) and an ionized gas sound speed of $c_{s,H\alpha} \sim 36$ km s$^{-1}$ (which is suitable for gas with temperatures of $T \sim 10^{5}$ K).  If we use instead the H$\alpha$ velocity dispersion in the cavity as representative of the flow speed, this would increase to $\mathcal{M} \sim 2.7 - 5.3$.  The molecular gas flow velocities in the shell (Fig. \ref{fig:hamom}) imply Mach numbers as high as $\mathcal{M} \sim 60$ in the shell (using a sound speed of $c_{s,CO} = 0.55$ km s$^{-1}$ suitable for $T \sim 10^{2}$ K gas).

Following \cite{Thompson16b}, the mass fraction of gas ejected from the system is then:
\begin{equation}
    \xi_{-}(x_{crit}) = \frac{1}{2}\left[1 - {\rm erf}\left(\frac{-2x_{crit} + \sigma_{{\rm ln} \Sigma}^{2}}{2\sqrt{2}\sigma_{{\rm ln} \Sigma}^{2}}\right)\right]
\end{equation}
where $x_{crit}$ describes the critical density susceptible to outflow and is related to the Eddington ratio as  $x_{crit} = {\rm ln}(\Gamma)$, and $\sigma_{{\rm} ln \Sigma}^{2}\simeq {\rm ln}(1 + R\mathcal{M}^{2}/4)$.  $R$ characterizes the ratio of variance in the ISM column densities to the volumetric densities - which is expected to vary depending on the exact form of the power spectrum of the gaseous ISM.  The authours note that for an assumption of isotropic and homogenous turbulence, a power-law density power spectrum ($P(k) \propto k^{-\alpha}$) is typical, in which case the value of $R$ depends on the Mach number as:
\begin{equation}
    R = \frac{1}{2}\left(\frac{3 - \alpha}{2-\alpha}\right)\left[\frac{1 - \mathcal{M}^{2(2-\alpha)}}{1-\mathcal{M}^{2(3-\alpha)}}\right]
\end{equation}
 In the last equation the authours suggest $\alpha = 3.7$ for $\mathcal{M}$ of a few, and $\alpha = 2.5$ for $\mathcal{M} \gg 1$.  In order to fully cover the expected final values permitted for $\xi$, we self-consistently compute R for the appropriate Mach numbers and $\alpha$ values spanned by the observed molecular and ionized components of the outflow.  This results in variations of $0.008 \leq R \leq 0.308$ and $0.245 \leq \sigma_{{\rm ln} \Sigma}^{2} \leq 2.104$.

From these parameter ranges we estimate permitted values for the fraction of mass escaping the system of $6\times 10^{-6} \leq \varsigma(\Gamma) \leq 6\times 10^{-3}$. For a star formation efficiency per free-fall time of $\epsilon_{\rm ff} = 0.01$ \citep{Krumholz14}, the implied mass loading factor produced by this theoretical estimate is:
\begin{equation}
0.0006 \leq \eta \equiv \varsigma/\epsilon_{\rm ff} \leq 0.6.
\end{equation}

This independent constraint on $\eta$ is an upper limit, as likely not all of the measured SFR couples energetically to the gas. Additionally the short conduction time suggests that the temperature structure within the inner part of the cavity will be homogenized efficiently ($\sim 10^{6}$ yr), perhaps inhibiting either cooling or condensation.  Despite these, it is already \textit{below} the critical mass-loading factor necessary to aid radiative cooling from the analysis in Section 6.2 ($\eta_{\rm crit} \geq 2.5$).  This then might suggest that a radiative cooling origin for the cold gas is plausible in general, but perhaps not occurring in NGC~3351 given the low SFR column densities and short travel time and distance.  

This is confirmed if we directly ask what fraction of the H$\alpha$ \textit{flux} in our MUSE observations is moving faster than the local escape velocity (as derived from our stellar dynamical modelling and independently confirmed via the CO circular velocity profile in \S 2.4). We find from the observations that only $5\times10^{-5}$ of the total H$\alpha$ flux is moving faster than the local escape velocity. This corresponds to an escaping \textit{mass} fraction of $3.7\times10^{-4} \leq \varsigma \leq 3.7\times10^{-3}$ given the range of electron densities in the cavity ($n_{e} \sim 10^{2-3}$).

Again assuming a star formation efficiency per free-fall time of $\epsilon_{\rm ff} = 0.01$ this produces an independent observational estimate of the initial mass loading factor of of:
\begin{equation}
    0.037 \leq \eta \leq 0.37
\end{equation}
These values are not only consistent with the estimates of the initial mass loading factor from our bubble expansion models, but again below the critical mass loading factor necessary for radiative cooling to play a strong role ($\eta_{crit} \lesssim 3$).
\newline

\subsection{Magnetic Entrainment}
Finally we consider the survival of the molecular gas in the context of our favoured dynamical scenario (\S 1.1) -- that the molecular gas was launched from the central ring primarily by direct photon pressure operating on the dust.  The molecular gas can be aided in its survival during such an outflow, by the galaxy's magnetic field lines \citep{Mccourt15}.  These would initially be wound through the dusty ISM in the ring, and could prevent the molecular shell from disrupting as it is pushed out, while also inhibiting conductive heating.

To further test the feasibility of this scenario we follow the arguments of \cite{Mccourt15}, who noted that the molecular shell feels a drag force as it expands outwards, due to the magnetic field lines -- akin to pushing against a net.  The magnetic field lines in the molecular shell would therefore impart a radius of curvature to the outflow front of:
\begin{equation}
R_{\rm curve} \sim \left(\frac{V_{\rm Alfv\acute{e}n}}{V_{\rm outflow}}\right)^{2}R_{\rm cloud}
\end{equation}

From the HST and ALMA images, the transverse dust and molecular gas shell has an observed radius of curvature which we measure to be $R_{\rm curve} = 340$ pc.  Using the measured outflow velocity from our H$\alpha$ gas, and a typical width to the shell of $R_{\rm cloud} = 40$ pc, we find the Alfv\'{e}nic velocity to be $V_{\rm Alfv\acute{e}n} = 235$ km s$^{-1}$.  With this we can solve for the magnetic field strength necessary to impart such a radius of curvature in the expanding molecular gas shell:
\begin{equation}
B = V_{\rm Alfv\acute{e}n}(4\pi\rho)^{1/2} = 330 \pm 20 \mu{\rm G}
\end{equation}
We note that the necessary magnetic field strength in this scenario is also consistent with the values of the fast shock models which reproduce our emission line ratios in Fig. \ref{fig:bpt}.

For comparison \cite{Thompson06} argue that galaxies should have an equilibrium magnetic field value which can help support disk self-gravity, which at the location of the ring in NGC~3351 would correspond to:
\begin{equation}
B_{\rm eq} \sim (8\pi^{2}G)^{1/2}\Sigma_{\rm gas} = 324 \mu{\rm G}
\end{equation}
This is in excellent agreement with our geometrically and dynamically derived value.  There are no direct observations of the magnetic field strength in NGC~3351, however indirect inferences from \cite{Thompson06} suggested $B_{\rm min} \geq 50 \mu{\rm G}$.  However, these authors argue strongly that in starburst galaxies the ambient magnetic field is likely always larger than this minimum.  As an upper bound, Eq. 1 of \cite{Lopez17} suggests $B_{\rm max} \leq 345 \mu{\rm G}$, which with $B_{\rm min}$ above, nicely bounds the geometrically derived $B$, and the theoretically motivated $B_{\rm eq}$ -- though consistent with the uncertainties.  

\cite{Mccourt15} also note an indirect estimate of the stopping distance for the wind in this magnetic entrainment scenario can be computed as: $d_{stall} \sim R_{cloud}(\rho_{cloud}/\rho_{wind})(\beta_{wind}\mathcal{M}^{2})/(1 + \beta_{wind}\mathcal{M}^{2})$.  The relative Mach number of the cloud to the wind ranges from $1.9 - 5.3$ for our flow, and from the best fit expansion models and Equation 21 $\beta_{wind} \sim 0.003$  Using these along with the typical densities in our CO cloud and wind, we find $d_{stall} \sim 150 - 1032$ pc.  The range of these expected stalling distances spans the observed present day travel distance of the CO shell $(R_{CO} \sim 520$ pc), however the large uncertainties only let us speculate that if the shell were indeed stalled, this may not be problematic for the magnetic entrainment scenario.  It should be kept in mind that there are significant systematic uncertainties in these simple analytic predictions, due to geometry or other factors, which prevent a conclusive statement on the validity of this scenario.  

Based on these simple calculations, a scenario where the cold gas/dust was aided in survival during its feedback driven outflow from the central ring by entrained magnetic fields appears possible.  This is likely not a unique scenario, nor does it need to operate singularly.  However this picture is consistent with our observations and energetic modeling of the outflow history (\S 3), and would reconcile the short conduction timescales ($t_{\rm cond} \sim 10^{6}$ yr) -- which derived in the absence of magnetic fields, may be significantly less than in the presence of strong field lines \citep{Thompson16b}.

The necessary magnetic field strengths (and orientation) for this scenario could perhaps be tested on this galaxy directly (or other comparable systems) with upcoming high resolution cm wavelength arrays, such as SKA -- and indeed VLA and SOFIA-HAWC+ observations of magnetic fields have already been demonstrated in several external galaxies (c.f., \citealt{Beck15}).

As a final summary, Table 1 shows several observed and derived relevant timescales for the primary outflow feature in NGC~3351.
 \begin{table}
 	\centering
 	\caption{Relevant gas timescales at the outflow front.}
% 	Remember to define the quantities, symbols and units used.}
 	\label{tab:example_table}
 	\begin{tabular}{lcl} % four columns, alignment for each
 		\hline
 		Timescale & Value & Notes\\
 		\hline
        \hline
	Outflow Travel & & \S 3.2, Appendix B\\
    log $t_{\rm exp}$ & $7.2_{-0.6}^{+0.6}$ & \\
    \hline
    Turbulent Dissipation & & Eq. 7\\
    log $t_{\rm diss}$ & $6.9_{-0.1}^{+0.2}$ & \\
    \hline
    Crushing & & \protect{\cite{Mccourt18}}\\
    log $t_{\rm crush}$ & $6.3_{-0.6}^{+0.4}$ & \\
    \hline
    Entrainment & & \protect{\cite{Mccourt15}}\\
    log $t_{\rm entr}$ & $6.9_{-1.0}^{+0.7}$ & \\
    \hline
    Cooling & & Eq. 8\\
    log $t_{\rm cool}$ & $7.9_{-0.8}^{+2.0}$ & for $\eta = 0.85 \pm 0.6$\\
    \hline
    Advection & & Eq. 9\\
    log $t_{\rm adv}$ & $5.8_{-0.4}^{+0.2}$ & for $\eta = 0.85 \pm 0.6$\\
    \hline
    Conduction & & \protect{\cite{Thompson16b}}\\
    log $t_{\rm cond}$ & $6.4$ & excluding magnetic fields \\
    \hline
 	\end{tabular}
 \end{table}

\subsection{Geometry and fate of the outflow}
\label{sec:geom}
At all radii the velocity dispersion of the ionised gas falls below the escape velocity of the galaxy, as derived from our JAM models and CO circular velocity curves; only a small fraction ($\sim 10^{-4}$) is escaping the total potential of the galaxy.

As noted by \cite{Swartz06}, due to the inclination of NGC~3351 it is uncertain as to whether the outflow (observed in their study as X-ray emission in the cavity) is expanding spherically or within the plane.  While naively one would expect expansion to be easier in the vertical direction, they suggest the outflow could be bounded in the plane of the disk -- perhaps confined by ambient gas in a halo leftover from previous outflow events -- though they could not fully rule out the possibility that the outflow is expanding vertically into the halo.  %We note that multiple outflow events would be consistent with the pattern of additional transverse shells seen to the north of the primary shell (Figures \ref{fig:hst3col} and \ref{fig:sfh}).

\cite{Swartz06} did suggest that the planar component of the outflow is indeed confined radially by the transverse dust shell, as the X-ray emission is bounded by this feature -- similar to the H$\alpha$ emission we observe.  They further note that the unusual orientation of the dust lane with respect to the bar lane means that it could have been generated during the outflow event -- consistent with the dynamical evidence presented here.

There are various analytic estimates of when a wind-driven outflow may breakout vertically from the plane of a galaxy.  Using for example, the parameterization from the study of \cite{MaclowMccray88}, we estimate 
the product of the mechanical luminosity and thermalization efficiency, $\xi\dot{E}$, to be a factor of $\sim 3-30$ above their estimated threshold for breakout.  This calculation assumes a isotropic outflow in a planar disk with scale height of $H = 100$ pc, and uses the luminosities and photoionisation pressures from the SB99 and MUSE emission line diagnostics in the cavity.  However the question of vertical confinement of gas is sensitive to the assumed scale height -- for example a change to $H = 200$ pc results in no breakout.  We note that assuming hydrostatic disk equilibrium ($H = \sqrt{2}R\sigma_{CO}/V_{CO}$) with our observed CO velocity dispersion favours $H \sim 200$ pc at the current location of the shell -- perhaps indicating the outflow has not expanded far out of the disk.

A vertically confined geometry is supported by considering the location and shape of the H$\alpha$ velocity dispersion peak in the outflow, relative to the peak of the radial outflow velocity and CO/dust shell.  In Fig. \ref{fig:haco2} we can see: 1) a spatial offset between the velocity dispersion maximum and the leading edge of the outflow and 2) a similar curvature to the high velocity dispersion ionised gas, as that of the CO shell.  Both provide clues to the geometry of the outflow.  For an optically thin gas which is undergoing expansion from the central ring, we can ask where the maximum value in the projected line of sight velocity dispersion should be for: a) the case of a spherically expanding ionised gas cavity, and b) a planar confined expansion.

Appendix E shows the relevant equations to model the line of sight velocity field and velocity dispersion.  For both geometries (spherical outflow, and vertically confined planar outflow), we assume an outflowing velocity field of $V(R) \propto R{\rm exp(}R/R_{\rm max}{\rm )}$, which is similar to the observed residual component of H$\alpha$ velocity in Fig. \ref{fig:hamom}.  The vertically confined outflow can be parameterized by a 3-D radius of curvature, which in turn sets a critical opening angle in the vertical ($\phi$) direction:
\begin{equation}
\phi_{\rm max} = {\rm arctan}\left(\frac{R_{\rm curve}}{R_{\rm max} - R_{\rm curve}}\right),
\end{equation}
where $R_{\rm max}$ is the projected maximum extent of the outflow on the sky.  We assume that the projected radius of curvature seen in the dust shell (\S 4.3.3) is representative of the 3-D $R_{\rm curve}$.  This yields an opening angle for the confined case of $\phi_{\rm max} = 29^{\circ}$. This would suggest a height above the disk of $h = 200\pm 50$ pc for the outflow at the location of the CO shell. 

\begin{figure}
\begin{center}
\includegraphics[width=0.48\textwidth]{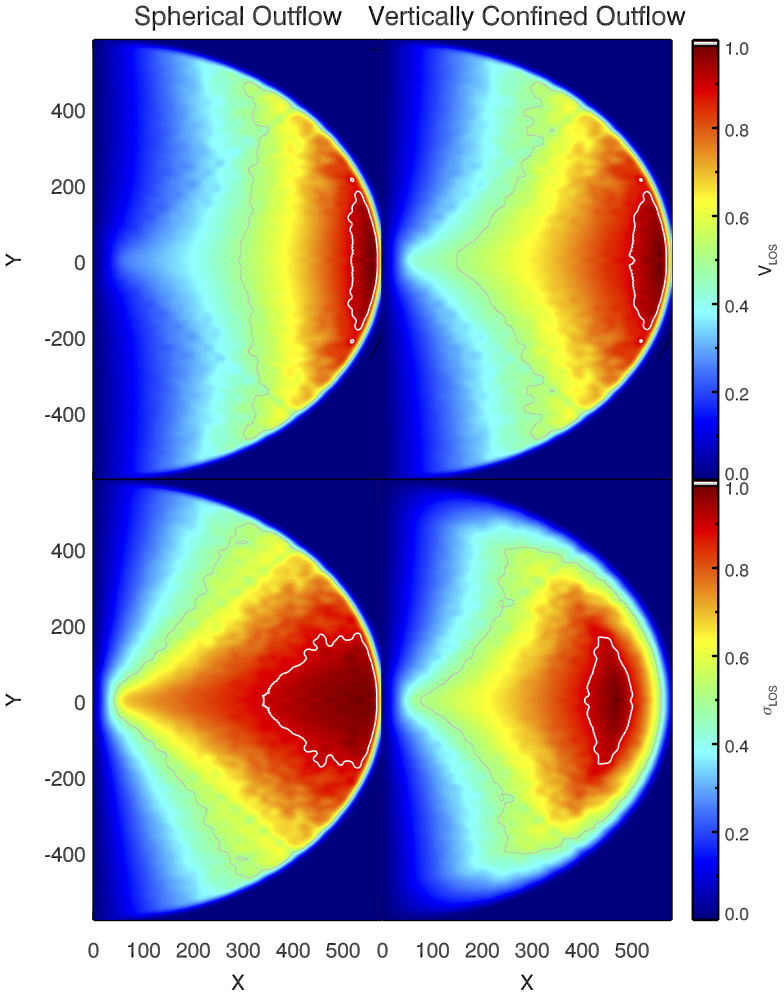}
\caption{\emph{Top row:} Radial component of the outflow velocity (projected on the sky) for an outflow that is spherically expanding (\emph{left}), or vertically confined (\emph{right}).\emph{Bottom Row:} The line of sight H$\alpha$ velocity dispersion in the two cases.  The confined outflow model shows an peak in the velocity dispersion offset from the leading edge, in close agreement with our observed kinematics.  All velocities are shown relative to the maximum value in the model velocity field, with the grey and white contours indicating levels at $50\%$ and $90\%$ of the maximum.}
\label{fig:timerpvmod}
\end{center}
\end{figure}

Figure \ref{fig:timerpvmod} shows the projected LOS residual velocities, and the LOS H$\alpha$ velocity dispersion in the two outflow cases.  While both residual velocity fields are consistent with our observations (by construction), only the vertically confined outflow model produces a well defined peak in velocity dispersion \emph{behind} the leading front of the outflow -- in excellent agreement with the $\sim 150$ pc observed offset between the $\sigma_{H\alpha}$ peak and the CO shell in Fig. \ref{fig:haco2}. This gives support to the above analytic arguments for the outflow being partially confined in the vertical direction.

Unlike some other systems in literature (e.g., \citealt{Alatalo11}) which show outflow velocities greater than the escape velocity, NGC~3351 appears to have a small scale nuclear outflow which can be powered by SF feedback and remain confined to the galactic potential.

\section{Discussion}
The observed gas kinematics, feedback models and simulations all imply that the transverse CO shell is not gravitational in nature, and that it is energetically feasible to have been ejected from the nuclear ring by stellar feedback.  Subsequent sustained energy from SF in the ring could move the ionised low density gas outwards in the cavity left behind, which we observe clearly as a radial component in the H$\alpha$ residual velocity map, orthogonal to the underlying galaxy velocity field.

While the scenario we present for the central region of NGC~3351 seems well supported by the data and our modelling, the unique boundary conditions from the multi-phase ISM observations also offers an opportunity to address additional questions on galactic astrophysics.

\subsection{Implications for Feedback Prescriptions in Simulations}
A result from our modeling of the dust shell expansion is that SNe alone can not provide the necessary energy to move the bulk of the gas from the SF ring to its observed location.  Indeed when we repeat the analysis of Sect. 3.2 with no direct photon pressure added to the energetics, we find that the shell expansion stops at roughly $R_{\rm gal} = 550$ pc ($d_{\rm stall} = 330$ pc) for the same parameters. 

While direct photon pressure is favoured as an additional energy source, we recover poor constraints on $\kappa$ as a result of the density, metallicity and SFR dependence in the photon re-scattering parameterizations. This can be seen most clearly as an inflection in the $\rho - SFR$ covariance plot in Fig. \ref{fig:covar}, as well as null correlations between $\kappa$ and the other parameters in that figure.  This unfortunately prevents us from the useful exercise of specifying an optimal $\kappa$ to use as a function of an observables such as SFR.

Nevertheless, while $\kappa_{\rm IR}$ values favoured in our analysis may be difficult to directly compare to subgrid simulation prescriptions, it could be possible to gain insight to the differential behaviour of $\kappa_{\rm IR}$ by repeating the exercise in Sect. 3.2 for other galaxies with similar feedback driven gas flows.  In this way perhaps relative changes in $\kappa$ with respect to galaxy mass or peak SFR could be observationally studied.

The absolute values of $\kappa$ spanned in the posterior distribution are larger than some currently implemented simulation prescritions, however these values are potentially degenerate with an unconstrained level of fractal/inhomogeneous structure for the unresolved molecular medium.

In the meantime, given the complex systematics in how simulations implement radiation pressure, perhaps a better lever arm is to ask what type of simulated feedback can reproduce the observed expansion time, ionised outflow momenta and emission line diagnostics -- when using the same SFRs and gas densities as observed here?  Can the morphology and dynamics of NGC~3351 be reproduced for a tailored simulation with feedback -- and if so what prescriptions succeed or fail?

In particular, as magnetic entrainment may be important to the survival of the cold phase, the accounting of feedback simultaneously in the presence of magnetic fields should encourage new generations of simulations to incorporate magnetohydrodynamical processes.

\subsection{Further Questions}
If the scenario presented here is a correct description of the dynamical evolution of the gas outside the nuclear star forming ring in NGC~3351, several consequences and questions emerge.

It would be interesting to understand how many other nuclear ring systems show similar morphology in the cold gas and dust in order to better characterize if there is a particular metallicity threshold necessary to efficiently eject these gas shells.  Or alternatively characterize if there is a specific SFR density or cadence that is necessary to generate such features.

A census and characterization of such transverse features may be beneficial as they hold not only a record of the outflow and star formation burst timescales in the galaxy, but also should provide a simple way of estimating the bar pattern speed.

From a chemical enrichment point of view, such large scale radial gas motions may transport metals a significant distance in the inner regions of the galaxy (c.f., \citealt{Gibson13}).  Similarly, more energetic flows, or those occuring in less dense disks may break out in the vertical direction and transport metals via a ``galactic fountain'' scenario.  If discrete enough in time, both of these gas motions may also produce azimuthal variations in metallicity due to the bar and ring rotation -- relevant in light of some studies finding variations in the abundances of spiral (inter)arm regions of galaxies \citep{Ho17}

Understanding the dynamics and energetics associated with these non-gravitational features in the gas will hopefully lead to improvements in our understanding of stellar feedback's impact on the ISM in local galaxies and simulations.

\section{Summary and Conclusions}
We have used MUSE, ALMA and HST observations of a transverse dust and molecular gas shell, located at the leading edge of an ionised gaseous outflow emanating from the central star-forming galactic ring in NGC~3351, to self-consistently describe its morphology and dynamics due to observable stellar feedback. Our key results are as follows:\\

\begin{itemize}
\item By running hydrodynamic gas response simulations with the derived gravitational potential of NGC~3351, we see that the observed transverse dust lane is likely not due to the underlying non-axisymmetric gravitational potential. As such, additional processes related to SF and stellar feedback are necessary to explain the presence and location of the transverse CO gas shell.

\item From the observed SFRs in the nuclear ring, we model the total momentum flux incident on the molecular gas shell using \texttt{STARBURST99} models and analytic feedback prescriptions.  We find that based on the total momentum flux incident on the molecular gas shell, the dominant source of momentum injection is from direct photon pressure ($\sim 40-60\%$), with SNe ($\sim 20-30\%$), Photoionisation heating ($\sim 10-15\%$), and stellar winds ($\sim 1-10\%$) progressively less important.

\item Dynamical modeling of the outflow, suggests the dusty, cold molecular gas was initially driven from the nuclear ring by SNe and radiation pressure, with an initial mass loading factor of $\eta \sim 0.1$, and took approximately $10^{7}$ years to reach its current location. This expansion time is close to the turbulent dissipation time of the molecular gas in the transverse shell at present.  

\item Concurrent star formation powered strong radial outflows of low density ionised gas, which we observe in the MUSE $H\alpha$ residual velocity field to have $V_{rad} \sim 70$ km s$^{-1}$, and cross orthogonally to the underlying circular orbits of the galactic velocity field.  This, and their confinement by the molecular gas shell, leads to visible shocks in the warm cavity bounded by the nuclear ring and dust shell.  The emission line diagnostics and velocities in this cavity are in agreement with predictions from fast-shock models.  

\item The geometry of the dust and ionised gas outflow is consistent with magnetic fields providing a (protective) drag force on the molecular gas in the outflow during its initial ejection by feedback from the nuclear ring.  \textit{In-situ} condensation and radiative cooling of the molecular gas are less likey for this system, as implied by the short dynamical timescale and low mass-loading factors.

%\item The short dynamical timescales and low mass-loading factors involved, suggest that neither in-situ condensation, nor radiative cooling are likely scenarios for the presence of CO at the leading front of the warm outflow.  Instead, the curvature of the dust shell is consistent with a magnetic drag force operating on the cold phase as it was launched out from the nuclear ring by stellar feedback.  This magnetic field could have helped the CO gas shell avoid conductive heating and disruption during the outflow.  The geometrically derived magnetic field strength is in excellent agreement with indirect observational constraints in this galaxy.

\item Comparison of the H$\alpha$ kinematics to the galaxy escape velocity profile, indicates that the outflowing gas will not escape the galaxy potential.

\item The morphology of the ionised gas velocity dispersion field relative to the outflow front, is most consistent with a vertically confined outflow, with an opening angle of $\sim 30^{\circ}$, suggesting it has not significantly broken out of the disk plane.   

\item The corresponding power generated by the star formation may be as high as $5\times10^{38}$ erg s$^{-1}$. This is of order of half that found in low-luminosity AGN (see \citealt{Cheung16}, and references therein). This provides at least one example that stellar feedback processes can imprint significant amounts of energy into the ISM even in galaxies showing no AGN activity.  Stellar feedback can likely to be an important driver of gas motions in high redshift systems where the SFR is larger than local systems, regardless of whether they have an AGN.
\end{itemize}
%5) The SFHs derived from spectra in separate azimuthal regions of the ring show differential changes which can be directly associated to discrete ionised gas outflow events and dust shells.

%6) The relative radial extent of consecutive dust shells provides a measure of the burst cadence of SF in the central ring, which is independent of pattern rotation in the galaxy.  From the four identified shells, we identify gaps between SF episodes ranging from $\Delta t_{burst} = 70-190$ Myrs.

%7) The angular separation of these multiple dust shells provides a novel geometric estimate of the bar pattern speed, which for constant ejection velocity of the molecular gas shells, implies $\Omega_{bar} = 37 \pm 5$ km s$^{-1} kpc^{-1}$.\\

  NGC~3351 shows quantifiable evidence of SF impacting the bulk (cold) gas motions in the disk plane, which has consequences for understanding outflows and feedback in galaxies.

If such nuclear SF episodes produce Kpc scale molecular outflows even without AGN, it may become non-trivial to distinguish the relative contributions of the two sources if AGN are also present in partially resolved systems. If molecular gas can be kept cool due to magnetic fields in the ISM, how frequent is this also in cold gas AGN-driven outflows?  The importance may be two-fold if the feedback momentum injection is also altered by the presence of magnetic fields.  These can both be addressed by continued advancements of MHD in simulations (e.g., \citealt{Pakmor17,Marinacci17}).

Observationally, an increased joint census of the molecular gas content in nearby starburst galaxies and direct measurements of their magnetic field strength with present and future cm wave interferometric facilities, will provide significant constraints for theoretical understanding of the interplay between feedback and magnetic fields.  Similar analyses as presented here, applied differentially to other galaxies with joint molecular and ionised gas observations, may lead to more helpful constraints for sub-grid prescriptions of stellar feedback in modern galaxy formation simulations, the survival and entrainment of cold gas in outflows, and a better understanding of the timescales for nuclear ring formation and evolution.

\section*{Acknowledgements}
We thank the anonymous referee for an extremely helpful referee report which greatly improved this manuscript.  The authors thank Morgan Fouesneau, Eva Schinnerer, Trevor Mendel, Thorsten Naab, I-Ting Ho, Rebecca McElroy, Alessandra Mastrobuono-Battisti, Chiara Mazzucchelli, Sebastien Viaene, Mattia Sormani and Naomi McClure-Griffiths for useful discussions which helped improve this manuscript. This work was supported by Sonderforschungsbereich SFB 881 ``The Milky Way System'' (subproject A7 and A8) of the Deutsche Forschungsgemeinschaft (DFG), and was also made possible with funding from the Natural Sciences and Engineering Research Council of Canada PDF award.  RL thanks AH. Based on observations collected at the European Organisation for Astronomical Research in the Southern Hemisphere under ESO programme 097.B-0640(A). RL, GvdV, AdLC and JF-B acknowledge support from grant AYA2016-77237-C3-1-P from the Spanish Ministry of Economy and Competitiveness (MINECO). PSB acknowledges support from grant AYA2016-77237-C3- 2-P. GvdV acknowledges funding from the European Research Council (ERC) under the European Union's Horizon 2020 research and innovation programme under grant agreement No 724857 (Consolidator Grant ArcheoDyn).

%%%%%%%%%%%%%%%%%%%%%%%%%%%%%%%%%%%%%%%%%%%%%%%%%%



\bibliographystyle{mnras}
\bibliography{timerblooie} % if your bibtex file is called example.bib

% \begin{thebibliography}{99}
% \bibitem[\protect\citeauthoryear{Author}{2012}]{Author2012}
% Author A.~N., 2013, Journal of Improbable Astronomy, 1, 1
% \bibitem[\protect\citeauthoryear{Others}{2013}]{Others2013}
% Others S., 2012, Journal of Interesting Stuff, 17, 198
% \end{thebibliography}

%%%%%%%%%%%%%%%%%%%%%%%%%%%%%%%%%%%%%%%%%%%%%%%%%%

%%%%%%%%%%%%%%%%% APPENDICES %%%%%%%%%%%%%%%%%%%%%

\appendix

\section{Simulation sub-grid feedback prescriptions}
While the SNe momentum flux, and wind momentum flux are directly output from the SB99 models, the direct photon-pressure momentum flux, depends not only on that output ionizing luminosity, but also on a prescription for how the photons may interact, re-scatter or become trapped within the dusty cold gas.

The baseline models for radiation pressure we adopt in the MCMC analysis of NGC~3351, come from Hopkins et al. 2014, 2017, which parameterize the opacity and (possible extra scattering) with a density and metallicity dependent factor - whereby more dense and metal rich gas should more effectively harness the momentum from the ionizing photons.

To understand systematics on this, Fig. \ref{fig:sbsys} illustrates systematic differences in the sub-grid prescriptions for radiative pressure used in several high resolution hydrodynamical galaxy formation simulations, which we describe below.

\subsection{Hopkins et al.}
The baseline model we adopt from Hopkins et al. (2014) allows for an effective absorption of the bolometric luminosity in high metallicity and density regimes, but does not require multiple re-scattering events per photon, and is given by:
\begin{equation}
\dot{p}_{\rm rad} = (1 - {\rm exp}(-\tau_{V}))(1 + \Sigma_{\rm gas}\kappa_{\rm IR})\frac{L}{c}
\end{equation}
where $\Sigma_{\rm gas}$ is the cold gas surface density, $\tau_{V}$ the optical depth in the V band, and $\kappa_{IR}$ a metallicity dependent mean IR dust opacity coefficient here taken as: $\kappa_{\rm IR} = 10 (Z/Z_{sun})$.

\subsection{Rosdahl et al.}
\cite{Rosdahl15} follow a similar formalism to Hopkins et al., but with an opacity factor $\kappa$, which varies as a function of wavelength from $\kappa_{\rm IR} = 10 (Z/Z_{sun})$ to $\kappa_{\rm opt,UV} = 1000 (Z/Z_{sun})$.  To illustrate the quantitative effect of the variation in $\kappa$, we show the difference in expansion timescales for the dusty CO shell in Fig. \ref{fig:key}.

\subsection{Agertz et al.}
\cite{Agertz13} allow for both a variation in the opacity and re-scattering coefficient, depending on if the radiation pressure acts on a timescale larger or shorter than a typical self-bound molecular cloud.  As seen in Fig. \ref{fig:sbsys}, the transition between these two regimes, nicely bounds the baseline models from \cite{Hopkins14} and \cite{Rosdahl15}.

The baseline model of \cite{Agertz13} that we adopt has the parameters: $\kappa_{IR} = 5 (Z/Z_{sun})$, $\mu_{\rm max} = 1$, $t_{\rm cloud} = 3$ Myr, $\eta_{1} = \eta_{2} = 2$, $\alpha = 0$, $\beta = 1.7$, $C_{R} = 2.5$, $M_{\rm min} = 100 M_{\odot}$, and $\epsilon_{\rm cloud} = 0.2$.  We run these assuming the maximum stellar mass (from the S4G mass maps; \citep{Querejeta15}) in the ring is $M_{max} = 10^{7} M_{\odot}$.

The final radiation pressure momentum flux in their model is:
\[
\dot{p}=\begin{cases}
               \dot{p}_{1} + \dot{p}_{2}  ~~~~~~~~~~~~~~~~~~~~~~; t < t_{cloud}\\
               (\eta_{1} + \eta_{2}\kappa_{IR}\Sigma_{gas})\frac{L}{c}   ~~~~~~~; t > t_{cloud}
            \end{cases}
\]
where,
\begin{eqnarray*}
 \dot{p}_{1} &=& \eta_{1}\frac{L}{c}\\
 \dot{p}_{2} &=& \frac{\eta_{2}\kappa_{\rm IR}(1-\epsilon_{\rm cloud})(2-\beta)}{2\pi C_{R}^{2}(3-2\alpha-\beta)}\left(\frac{\mu_{\rm max}}{\epsilon_{cloud}}\right)^{1-2\alpha} \times \\
             && \frac{1 - \left(\frac{M_{\rm min}}{M_{\rm max}}\right)^{3-2\alpha-\beta}}{1 - \left(\frac{M_{\rm min}}{M_{\rm max}}\right)^{2-\beta}}\frac{L}{c}
\end{eqnarray*}
             
\subsection{Krumholz et al.}
Here the effective increase in the direct photon pressure is specified through a trapping coefficient $f_{\rm trap}$, which enters in an analogous way to $\kappa_{\rm IR}$ in the above prescriptions.  The trapping factor is given a physical link by being expressed as a function of two quantities: the mean optical depth $\tau_{*}$ and the ratio of radiative to gravitational forces $f_{E}$.  Following the scalings suitable for disk systems in \cite{KrumThomp13} these parameters are set as:
\begin{eqnarray}
\tau_{*} = 0.67\frac{f_{\rm gas}}{0.5}\left(\frac{L/SFR}{10^{10}}\right)^{1/2}\left(\frac{\Sigma_{SFR}}{10^{3} M_{\odot} pc^{-2} Myr^{-1}}\right)^{1/2}\left(\frac{\Sigma_{\rm gas}}{g cm^{-2}}\right)\\
f_{E} = 0.43\left(\frac{L/SFR}{10^{10}}\right)^{3/2}\left(\frac{\Sigma_{\rm SFR}}{10^{3} M_{\odot} pc^{-2} Myr^{-1}}\right)^{3/2}\left(\frac{g cm^{-2}}{\Sigma_{\rm gas}}\right)
\end{eqnarray}

For NGC 3351, the observed luminosities, SFRs and gas densities give maximal values of ($\tau, f_{E}, f_{\rm trap}) \simeq (0.004,0.03,0.002)$, indicating we are still in a low SF regime for these models.

As noted by \cite{KrumThomp12}, even in higher SFR surface density regimes, the effective trapping is lower in their models than in the \cite{Hopkins14} prescriptions, which can be seen quantitatively for the application of the models to NGC~3351 in Fig. \ref{fig:sbsys}.

\section{Star formation history of NGC~3351's nuclear ring}
The MUSE spectra contain a wealth of information on the stellar populations within NGC 3351.  To recover the recent SFH in different regions of the central ring in NGC~3351 we use the full spectral fitting routine \texttt{STECKMAP} (STEllar Content and Kinematics via Maximum a Posteriori; \citealt{Steckmapa,Steckmapb}).  The justification and testing of this methodology is outlined in e.g., \cite{SanchezBlazquez11,RuizLara15}, and our adopted methodology is described in detail in \cite{SanchezBlazquez14}.  In short, the integrated spectrum within a given spaxel is compared to a grid of composite stellar population models in order to recover the light weighted mass fractions in different age bins.  This is a non-parameteric method for recovering the SFH and does not impose any particular shape for the solutions.

The SFH analysis utilizes the models from \cite{Vazdekis10}, which have been computed with the MILES stellar library \citep{SanchezBlazquez06,Cenarro07}. The SFH is inferred from the spectral fits, after first fixing the kinematic parameters ($\sigma, V_{sys}$) with pPXF \citep{ppxf}, correcting for instrument response and extinction effects to the stellar continuum, and masking emission lines and sky residuals.

Figure \ref{fig:sfh} shows the derived mass fractions as a function of time (SFH) in the last 120 Myr for three regions in the star forming central ring.  The left hand panel shows a map of the central region of NGC~3351, in which the primary dust shell is clearly seen (labeled S1), however in addition a smaller dust shell (S2) is also visible.  This secondary shell shows an associated kinematic outflow in the H$\alpha$ residual velocity field, stellar feedback from the ring has occured discretely in time, and powered multiple outflow events.  

%There is some evidence for this when considering that the SFH of the magenta region in Figure \ref{fig:sfh} near the primary outflow feature, shows the largest SFRs at oldest times.  Conversely the regions which have the strongest recent SFH are located incident to a smaller transverse dust feature (green region in Figure \ref{fig:sfh}).  %This dust substructure (labelled S2 in Figure \ref{fig:sfh}) shows a clear kinematic counterpart (V2), with an outflow feature in our residual H$\alpha$ velocity map that is clearly bounded by the smaller dust shell - analogous to the primary ionised gas outflow and main dust shell (S1 and V1).

%It is interesting to note that SFH appears to continue to rise as it approaches the time derived from our bubble expansion models ($\sim 3-9$ Myr).  
While the age limits ($t_{\rm min} = 30$ Myr) of the stellar library prevent a firm link between the recent SF of the ring and any morphological features, it appears likely that the peak SFR in the ring occurred sometime in the last 30 Myr, consistent with the dynamical analysis presented in the subsequent section. 

%One can however infer a differential timescale between the bursts that created the main dust shell, and the smaller sub-shell.  As their current maximum observed outflow velocity is similar ($V_{rad} \sim 70$ km s$^{-1}$), simple geometry gives an estimate for the timescale for the subsequent starbursts (independent of the radial distances of the two shells). 
% \begin{equation}
% \Delta t_{burst} = \frac{(r_{1} - r_{2})}{V_{ej}}.
% \end{equation}

% From their relative positions on the HST image, and a coarse estimate of the time integrated ejection velocity of $V_{ej} = V_{res} = 70$ km s$^{-1}$, we find $\Delta t_{burst} = 4.4 \pm 1.7$ Myrs.  This is much shorter than the local gas cooling time and a fraction of the orbital time, suggesting that cold molecular gas is introduced for SF along the ring from an external source - likely along the bar feeding arms, with SF triggered at the contact points with the ring (e.g., \citealt{vandeven09,Boker08}).  This will be discussed in more detail with respect to our numerical hydrodynamical simulations in the below section. We note the angular offset between the two shell features is also consistent with the location of incident SF in the ring moving due to rotation of the contact points at the pattern speed of the bar.  This will be explored more in a forthcoming work (Leaman et al. in prep).

%Measuring the radial extent of the four shell features, we can define three time intervals, for which we find peak SFR gaps of $\Delta t_{burst} \sim 70, 148, 187$ Myr.

\begin{figure}
\begin{center}
\includegraphics[width=0.48\textwidth]{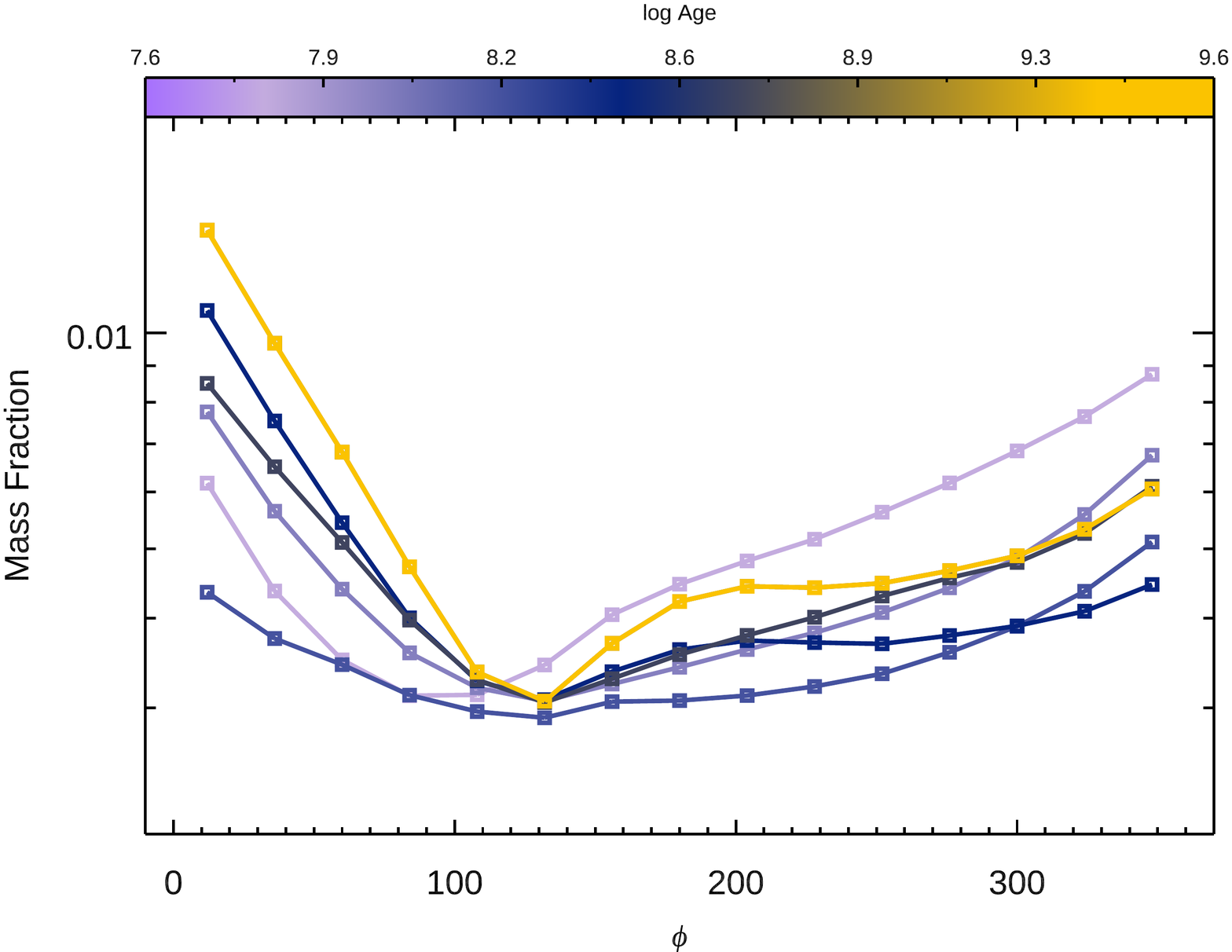}
\caption{Mass fraction in different age bins, as a function of azimuthal angle in the ring ($\phi = 0$ is aligned with North on the galaxy image).  There is perhaps a mild phase shift in the minimum for the youngest age bins, but as a whole, to first order the ring SFH appears relatively coherent on these timescales.}
\label{fig:timerpphase}
\end{center}
\end{figure}

The first order result from the SFH analysis of the ring is that there is a coherent SFH across the ring, at least on timescales of a few hundred Myr.  This can be seen in Fig. \ref{fig:timerpphase} which shows the SFR in different age bins as a function of azimuthal sector in the ring.  While there is a mild phase shift for the youngest age bin, the well matched curves would suggest that SFR variations in the ring are likely driven by the large scale fueling of the ring via the feeding arms.  This is perhaps not surprising if azimuthal variations were created (and erased) on the orbital timescale of stars in the ring ($\sim 10$ Myr), which is below the age resolution of our youngest SFH bin.

\begin{figure}
\begin{center}
\includegraphics[width=0.48\textwidth]{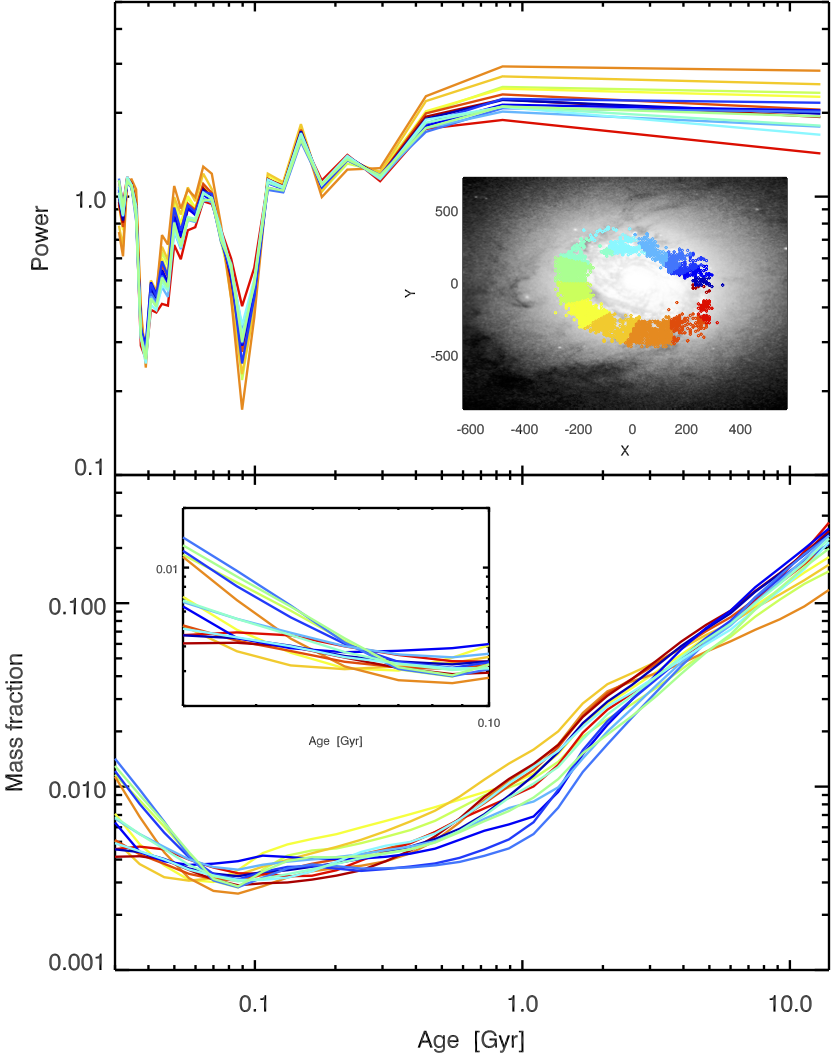}
\caption{Power spectrum of the SFHs in different azimuthal sectors in the ring.  There is a mild secondary effect with sectors nearest opposite contact points (orange, light blue) showing offset power at fixed period.  Bottom panel shows the SFHs for all sectors.}
\label{fig:timerppow}
\end{center}
\end{figure}

Smaller correlations amongst the various azimuthal sectors are visible as a very mild second order effect in Fig. \ref{fig:timerppow}, which shows the full SFHs and power spectrum of the SFHs as a function of azimuthal sector.  For example, the sectors at opposing contact points (cyan/light blue vs. orange) show consistently separated tracks in the power spectrum diagnostic, though the behaviour is mild, as indicated by the lack of visible trend in the bottom panel.

%We caution again that a direct link between spatial variations in the ring's SFH and outflow features may be complicated due to the fact that the orbital time in the ring is approximately 10 Myr and the rotation of the contact points along the ring due to the bar pattern speed should take place over $\sim 200$ Myr.

\section{Covariance Diagnostics for the Feedback Models}
\begin{figure}
\begin{center}
\includegraphics[width=0.48\textwidth]{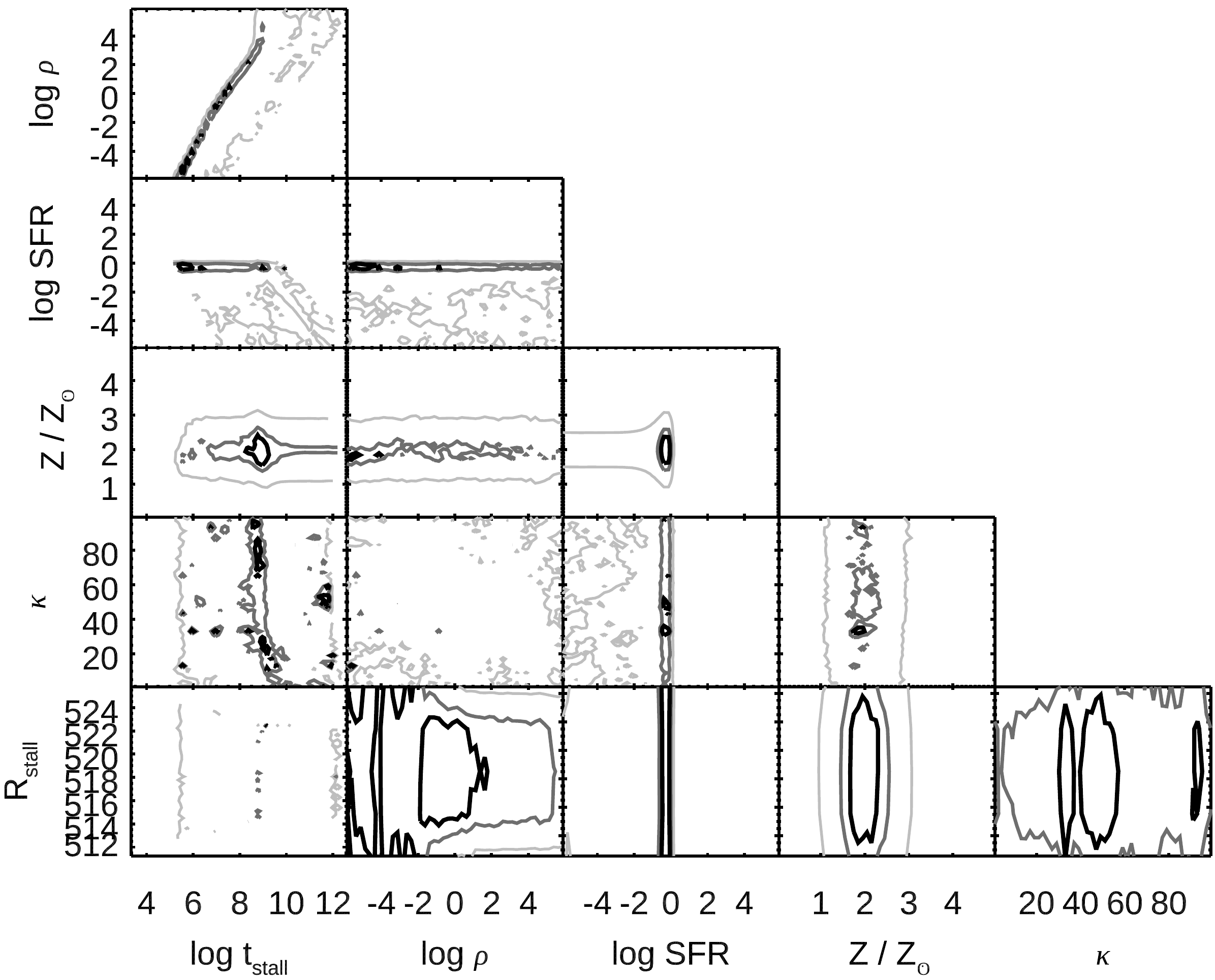}
\caption{Covariance diagrams for the shell expansion model parameters. The density and metallicity dominate the final expression for momentum flux in the radiation pressure expression we use (Appendix A) and this as a consequence prevents useful constraints on $\kappa$.}
\label{fig:covar}
\end{center}
\end{figure}

For completeness, Fig. \ref{fig:covar} shows the parameter correlations from the feedback model MCMC runs in Sect. 3.2.

\section{Model Velocity Field Setup}
To construct the line of sight velocity dispersion maps for an optically thin expanding outflow, we consider integration of the underlying 3-D velocity field along a sightline, in one of two underlying geometries: spherical outflow, or a vertically confined outflow.  The intrinsic velocity field is defined in both geometries as $V(R,\theta,\phi) = V_{0}R{\rm exp(}(R/R_{\rm shell}{\rm )}$, where $\theta$ is the azimuthal angle in the $X-Y$ plane and $\phi$ the polar angle.

\subsection{Spherical Outflow}
\begin{figure}
\begin{center}
\includegraphics[width=0.48\textwidth]{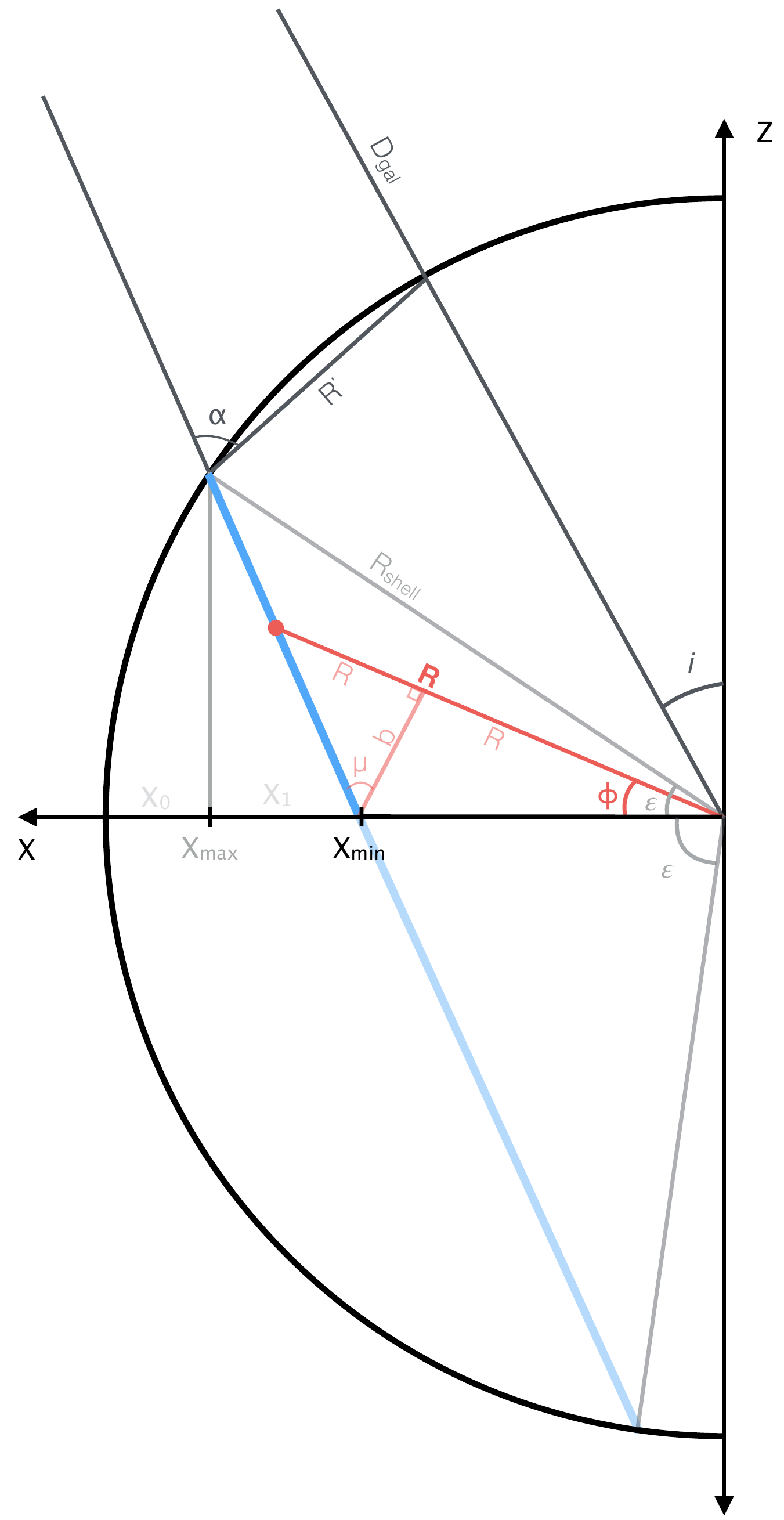}
\caption{Geometry of line-of-sight through a spherical outflow of optically thin gas, in a galaxy seen at inclination $i$, here shown in the X-Z plane.}
\label{fig:vfieldsph}
\end{center}
\end{figure}

Figure \ref{fig:vfieldsph} shows the geometry of an arbitrary sightline which strikes the spherical outflow at a projected galactocentric distance of $R^{'}$, in a galaxy that is at distance $D_{\rm gal}$ from the observer, seen at inclination $i$.  

The line of sight can be parameterized by an angle from the plane of the galaxy $\epsilon$.  To parameterize the location along the line of sight through the velocity field (blue segment in Fig. \ref{fig:vfieldsph}), we consider a vector from the origin to a distance $R$, with angle $\phi$.  As $\phi$ sweeps from $0 \leq \phi \leq \epsilon$, $R$ corresponds to the 3-D distance from the origin at that point on the line of sight segment.

The line of sight velocity distribution at a projected distance $R^{'}$ and planar azimuthal angle $\theta$ in the galaxy then becomes:
\begin{equation}
\int_{\epsilon=0}^{\epsilon=90-i} \int_{\phi=\epsilon_{-}}^{\phi=\epsilon} V_{LOS}(R,\theta) d\phi d\epsilon
\end{equation}

An arbitrary sightline (parameterized by $\epsilon$) will enter the spherical shell at an angle $\alpha = 90 - {\rm tan^{-1}}(R^{'}/D_{\rm gal})$.  This is related to $\epsilon$ by
\begin{equation}
\epsilon = 180 - \alpha -i
\end{equation}

As we are considering an optically thin gas, and due to the intermediate inclination of our galaxy ($i \sim 45^{\circ}$), the line of sight will exit the shell asymmetically at a location corresponding to a polar angle below the plane $\epsilon_{-}$, defined by:
\begin{eqnarray}
\epsilon_{-} &=& {\rm arccos}\left(\frac{X_{-}}{R_{shell}}\right);\\
X_{-} &=& X_{min} - \beta{\rm cos}(\epsilon)\\
\beta &=& R_{shell}\left(2{\rm sin}(\alpha) - \frac{{\rm sin}(\upsilon)}{{\rm sin}(\epsilon)}\right)\\
X_{min} &=& R_{shell}\left({\rm cos}(i+\upsilon) - \frac{{\rm sin}(i+\upsilon)}{{\rm tan}(\epsilon)}\right)
\end{eqnarray}
Here $\upsilon = 90 - (i+{\rm arcsin}(R^{'}/R_{shell}))$.

Finally the 3-D radius at a position on the line of sight segment is given by:
\begin{eqnarray}
R &=& R_{shell}\left({\rm cos}(90-\upsilon) - \frac{{\rm sin}(90-\upsilon)}{{\rm tan}(\epsilon)}\right)\times \\
&&\large({\rm sin}(90-\phi){\rm tan}(\alpha-i+\phi) \pm {\rm cos}(90-\phi)\large)\nonumber
\end{eqnarray}

The sign of the last argument is negative for the portion of the line of sight segment above the $Z$ plane of the galaxy (e.g., $0 \leq \phi \leq \epsilon$), and positive for the portion corresponding to $\epsilon_{-} \leq \phi \leq 0$.

For a given line of sight (parameterized by $\theta$), the deprojected velocity at a 3D radius $R$ (parameterized by $\phi$) $V(R)$ can then be transformed into a LOS component:
\begin{equation}
V_{LOS}(R,\phi,\theta) = \left(\frac{\pm V(R)}{{\rm sin}(\phi)}\right){\rm cos}(\theta)
\end{equation}

The sign of the term will change depending on if $\phi$ is above or below the plane of the galaxy.  With this, the projected line-of-sight velocity dispersion is computed as the second moment of the integral in Eq. E1.

\subsection{Vertically Confined Outflow}
\begin{figure}
\begin{center}
\includegraphics[width=0.48\textwidth]{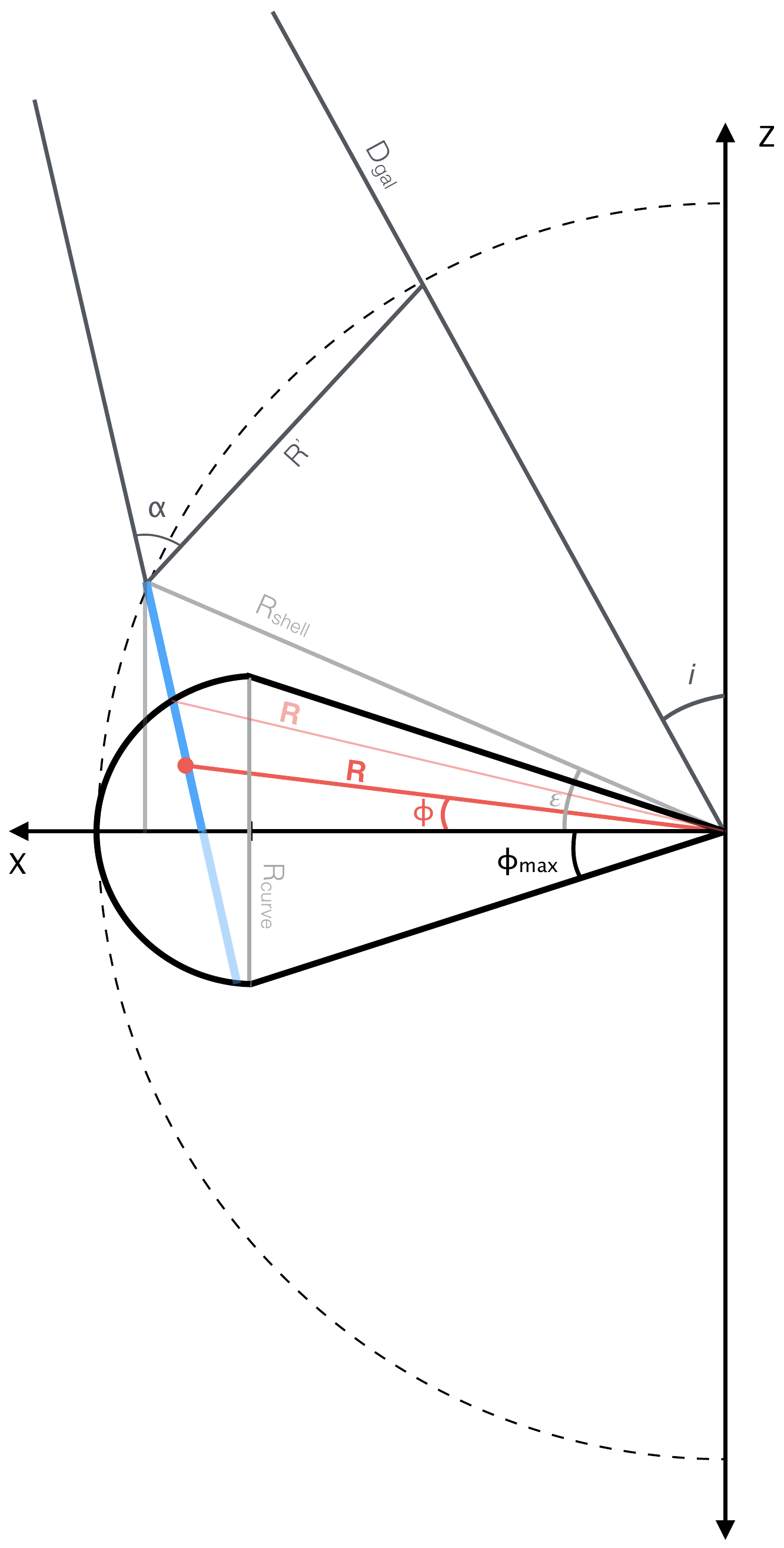}
\caption{Geometry of line-of-sight through a vertically confined outflow of optically thin gas, in a galaxy seen at inclination $i$, here shown in the X-Z plane.}
\label{fig:vfieldc}
\end{center}
\end{figure}

We next consider the case of an outflow confined in the vertical direction, with some critical opening angle $\phi_{\rm max} = {\rm arctan}(R_{\rm curve}/R_{\rm off})$, which is defined by a radius of curvature for the confined outflow, relative to the maximum radial extent of the outflow $R_{\rm off} = R_{\rm shell} - R_{\rm curve}$.  The geometry in the X-Z plane is similar to that of an ice-cream cone and is shown in Fig. \ref{fig:vfieldc}.  

The velocity field is constructed in the same way as in the spherical case, however outside of $\pm \phi_{\rm max}$ the velocity is zero.  This is trivial for lines of sight with $X_{max} \leq R_{\rm off}$ which pass through the ``waffle cone'' portion of the outflow.  However as $R_{\rm curve} \leq R_{\rm shell}$, for sightlines that pass through the curved ``ice cream'' portion of the outflow, the segment length will no longer be given by the same limits as in the spherical case. Instead the 3-D distance to a point along the line-of-sight segment, $R$, runs from $X_{min}$ to some maximal radius defined by the smaller radius of curvature of the outflow:
\begin{equation}
R_{\rm conf} = \frac{1}{2}\left(\sqrt{2R_{\rm off}^{2}{\rm cos}(2\phi)-R_{\rm off}^{2} + 2R_{\rm curve}^{2}} + 2R_{\rm off}{\rm cos}(\phi)\right)
\end{equation}
With this, the velocity field can be defined exclusively within the region of interest, and the LOS velocity distribution solved as in the spherical case using Eq. E1$-$E8.

%%%%%%%%%%%%%%%%%%%%%%%%%%%%%%%%%%%%%%%%%%%%%%%%%%

% Don't change these lines
\bsp	% typesetting comment
\label{lastpage}
\end{document}